\documentclass[aoas,nameyear,dvips]{arximspdf}
\usepackage{dcolumn,multirow}
\usepackage{subenv}
\usepackage{graphicx}

\doi{10.1214/09-AOAS305}
\volume{4}
\issue{2}
\pubyear{2010}
\firstpage{1105}
\lastpage{1138}

\makeatletter
\newcolumntype{d}[1]{D{.}{.}{#1}}
\makeatother

\begin{document}
\begin{frontmatter}

\title{Intervention analysis with state-space models to estimate
discontinuities due to a survey redesign\protect\thanksref{T1}}
\runtitle{Intervention analysis with state-space models}
\begin{aug}
\author[A]{\fnms{Jan} \snm{van den Brakel}\ead[label=e1]{jbrl@cbs.nl}\corref{}}
\and
\author[A]{\fnms{Joeri} \snm{Roels}}
\thankstext{T1}{The views expressed in this
paper are those of the authors and do not necessarily reflect the
policies of Statistics Netherlands.}
\runauthor{J. van den Brakel and J. Roels}
\affiliation{Statistics Netherlands}
\address[A]{
Department of Statistical Methods\\
Statistics Netherlands\\
P.O. Box 4481\\
6401 CZ Heerlen\\
The Netherlands\\
\printead{e1}}
%adresu isvedimo komanda gale!
\end{aug}

% HISTORY:
\received{\smonth{12} \syear{2008}}
\revised{\smonth{10} \syear{2009}}

% ABSTRACT
%
\begin{abstract}
An important quality aspect of official statistics produced by national
statistical institutes is comparability over time. To maintain
uninterrupted time series, surveys conducted by national statistical
institutes are often kept unchanged as long as possible. To improve the
quality or efficiency of a survey process, however, it remains
inevitable to adjust methods or redesign this process from time to time.
Adjustments in the survey process generally affect survey
characteristics such as response bias and therefore have a systematic
effect on the parameter estimates of a sample survey. Therefore, it is
important that the effects of a survey redesign on the estimated series
are explained and quantified. In this paper a structural time series
model is applied to estimate discontinuities in series of the Dutch
survey on social participation and environmental consciousness due to a
redesign of the underlying survey process.
\end{abstract}
%
% KEYWORDS
%

\end{frontmatter}

%s1 ###
\section{Introduction} \label{sec:1}

Surveys conducted by national statistical institutes are generally
conducted continuously or repeatedly in time with the purpose to produce
consistent series. Quality of official statistics is based on various
dimensions; see Brackstone (\citeyear{Brackstone99}) for a discussion. One important
quality aspect is comparability over time. To produce consistent series,
national statistical institutes generally keep their survey processes
unchanged as long as possible. It remains inevitable, however, to
redesign survey processes from time to time to improve the quality or
the efficiency of the underlying survey process. In an ideal survey
transition process, the systematic effects of the redesign are explained
and quantified in order to keep series consistent and preserve
comparability of the outcomes over time. There are various possibilities
to quantify the effect of a survey redesign; see van den Brakel, Smith
and Compton (\citeyear{vandenBrakelSmith08}) for an overview. If the redesign affects the data
collection phase, then a parallel run is a reliable approach to avoid
the confounding of real changes in the underlying phenomenon of interest
with the systematic effect of the redesign. Therefore, the redesign of
long-standing surveys like, for example, the US Current Population
Survey and the US National Crime Victimization Survey, are accompanied
with a parallel run [Dippo, Kostanich and Polivka (\citeyear{Dippo94}) and Kindermann and Lynch
(\citeyear{Kindermann97})].

Significance and power constraints necessary to establish the
prespecified treatment effects generally require large sample sizes for
both the regular and the new survey in the parallel run. This is not
always tenable due to budget constraints. The National Health Interview
Survey (NHIS), established in 1956, is another example of a long
standing survey. This survey was radically redesigned in 1997 [Fowler
(\citeyear{Fowler96})]. The absence of a parallel run obstructed the analysis of trends
in different key variables of the NHIS. Akinbami and Schoendorf (\citeyear{Akinbami02}) and
Akinbami, Schoendorf and Parker (\citeyear{Akinbam03})
reported that trends in estimates of childhood asthma prevalence are
disrupted due to changes in the NHIS design in 1997, which created the
impression that childhood asthma prevalence declined in this period.
Caban et al. (\citeyear{Caban05}) used NHIS data to study trends in prevalence rates
of obesity among working adults. Data were analyzed separately for NHIS
periods 1986 until 1995 and 1997 until 2002 because of the major
redesign of the NHIS in~1997. These examples illustrate that in
situations were no parallel run is available, alternative methods, which
are based on explicit statistical models, should be considered to
quantify the effect of a redesign. In this paper an intervention
analysis using structural time series models is proposed as an
alternative for conducting large scale field experiments and applied to
a real life example at Statistics Netherlands. This is a direct
application of the intervention approach proposed by Harvey and Durbin
(\citeyear{Harvey86}) to estimate the effect of seat belt legislation on British road
casualties.

In survey methodology, time series models are frequently applied to
develop estimates for periodic surveys. Blight and Scott (\citeyear{Blight73}) and
Scott and Smith (\citeyear{Scott74}) proposed to regard the unknown population
parameters as a realization of a stochastic process that can be
described with a time series model. This introduces relationships
between the estimated population parameters at different time points in
the case of nonoverlapping as well as overlapping samples. The explicit
modeling of this relationship between these survey estimates with a time
series model can be used to combine sample information observed in the
past to improve the precision of estimates obtained with periodic
surveys. This approach is frequently applied in the context of small
area estimation. Some key references to authors that applied the time
series approach to repeated survey data to improve the efficiency of
survey estimates are Scott, Smith and Jones (\citeyear{Scott77}), Tam (\citeyear{Tam87}), Binder and Dick
(\citeyear{Binder89,Binder90}), Bell and Hillmer (\citeyear{Bell90}), Tiller (\citeyear{Tiller92}), Rao and Yu (\citeyear{Rao94}),
Pfeffermann and Burck (\citeyear{Pfeffermann90}), Pfeffermann (\citeyear{Pfeffermann91}), Pfeffermann and Bleuer
(\citeyear{Pfeffermann93}), Pfeffermann, Feder and Signorelli (\citeyear{Pfeffermann98}), Pfeffermann and Tiller (\citeyear{Pfeffermann06}), Harvey
and Chung (\citeyear{Harvey00}), Feder (\citeyear{Feder01}) and Lind (\citeyear{Lind05}).

In 1997 Statistics Netherlands started the Permanent Survey on Living
Conditions (PSLC). This is a module-based integrated survey combining
various themes concerning living conditions and quality of life. Two
modules of the PSLC, the Module Justice and Environment and the Module
Justice and Participation, are used to publish figures about justice and
crime victimization. The first module is also used to publish figures
about environmental consciousness. The second module is used
additionally to publish information about social participation. To
realize expenditure cuts, the PSLC stopped at the end of 2004. From that
moment on, figures about social participation and environmental
consciousness are based on a separate survey, called the Dutch Survey on
Social Participation and Environmental Consciousness (SSPEC).

In this survey transition the data collection mode, the questionnaire,
the context of the survey and the fieldwork period changed, which
resulted in systematic effects in the outcomes of the survey. Since the
redesign mainly affects the data collection process in this application,
a large scale field experiment is very appropriate to test the effect on
the parameter estimates of the survey; see, for example, van den Brakel
(\citeyear{vandenBrakel08}). An experimental approach might, however, be hampered due to
budget and other practical constraints, which was the case for the Dutch
SSPEC. Therefore, an intervention analysis using a structural time
series model is used as an alternative to quantify the
effect of the
redesign on the main series of the sample survey.

All target variables of the PSLC and the SSPEC have multinomial
responses which are transformed to proportions of units classified in
$K\ge2$ categories. The survey estimates of these proportions are
observed on a $(K-1)$-dimensional simplex and comprise a composition.
Aitchison (\citeyear{Aitchison86}) developed statistical methods for the analysis of
compositional data, using additive logratio and central logratio
transformations. Brunsdon and Smith (\citeyear{Brunsdon98}) developed VARMA models for
logratio transformed compositional time series. Silva and Smith (\citeyear{Silva01})
applied the structural time series modeling approach to logratio
transformed compositional time series. In this paper the intervention
approach proposed by Harvey and Durbin~(\citeyear{Harvey86}) is applied to estimate the
effect of a survey redesign on compositional time series obtained with
periodic surveys.

In Section \ref{sec:2} the PSLC and the SSPEC are described. The systematic
effects due to the redesign are discussed in Section \ref{sec:3}. A time series
model to quantify these discontinuities is developed in Section \ref{sec:4}.
Results for the most important indicators for four different models are
given in Section \ref{sec:5}. The performance of these models are investigated in
a simulation study, which is also described in Section \ref{sec:5}. The paper
concludes with a discussion in Section \ref{sec:6}.

%s2 ###
\section{Survey designs} \label{sec:2}

%s2.1 ###
\subsection{Permanent survey on living conditions} \label{sec:2.1}

The PSLC was conducted as a repeatedly cross sectional survey, which
implies that there is no sample overlap in time. The Module Justice and
Environment and the Module Justice and Participation of the PSLC use
persons aged 15 years or older as the target population. The PSLC was a
continuously conducted survey. Each month a self-weighted stratified
two-stage sample of persons was drawn from a sample frame derived from
the municipal basic registration of population data. Strata are formed
by geographical regions. Municipalities are considered as primary
sampling units and persons as secondary sampling units. The monthly
sample size averaged between 550 and 700 persons for both modules. With
response rates varying around a level of 60\%, this resulted in a yearly
net response of about 4000 to 5000 persons for both modules.

Interviewers visited all the sampled persons at home and administered
the questionnaire in a face-to-face interview. This is generally
referred to as computer assisted personal interviewing (CAPI). The
estimation procedure used to compile official statistics is based on the
generalized regression estimator [S\"{a}rndal, Swensson and Wretman (\citeyear{Sarndal92}), Chapter 6]
using a weighting scheme that is based on different sociodemographic
categorical variables.

%s2.2 ###
\subsection{Survey on social participation and environmental
consciousness} \label{sec:2.2}

The PSLC stopped at the end of 2004. From that moment figures about
social participation and environmental consciousness are based on the
SSPEC. This survey is also conducted as a repeatedly cross sectional
survey and is based on a self-weighted stratified two-stage sample
design of persons aged 15 years and older residing in the Netherlands.
Data are collected by computer assisted telephone interviewing (CATI).
As a result, the subpopulation aged 15 years and older with an unlisted
telephone number or cell-phone number is not observed. The data
collection of the SSPEC is conducted in the months September, October
and November with a monthly sample size of about 2500 persons. The
estimation procedure is, like the PSLC, based on the generalized
regression estimator. The response rates in the SSPEC varied around
65\%. As a result, about 4500 respondents are observed in the yearly
samples.

Since 2005, figures about justice and crime victimization are based on
the Dutch Security Monitor. See van den Brakel, Smith and Compton (\citeyear{vandenBrakelSmith08})
for more details about this redesign and the effects on the main series
of this survey.

%s2.3 ###
\subsection{Target parameters} \label{sec:2.3}

All target variables about environmental consciousness and social
participation are based on closed questions where the respondent can
choose one out of $K$ answer categories to specify his opinion or
behavior on an ordinal scale. The target parameters are the estimated
proportions that specify the distribution over these $K$ categories for
the entire population or subpopulations. In this paper the series of two
variables are used for illustrative purposes. The first variable,
Separating chemical waste, is an example of environmental consciousness.
This variable contains five answer categories: (1)~always, (2)~often,
(3)~sometimes, (4)~rarely and (5)~never. The second variable, Contact
frequency with neighbors, is an example of social participation. This
variable contains four answer categories: (1) at least once a week, (2)
once within two weeks, (3) less than once within two weeks and (4)
never. An overview of all target variables can be found in the
supplemental paper, van den Brakel and Roels (\citeyear{vandenBrakelRoels09}).

%s3 ###
\section{Factors responsible for discontinuities} \label{sec:3}

\begin{figure}

\includegraphics{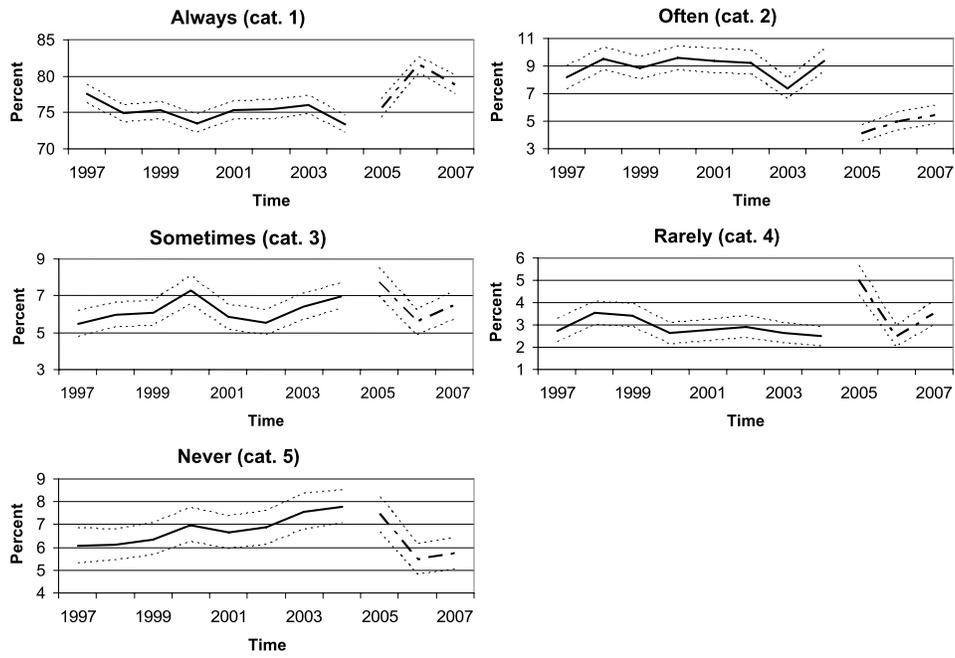}

\caption{Separating chemical waste.
Solid line: observed series under the PSLC, dashed line: observed series
under the SSPEC, dotted line: 95\% confidence interval.}\label{fig1}
\end{figure}

\begin{figure}

\includegraphics{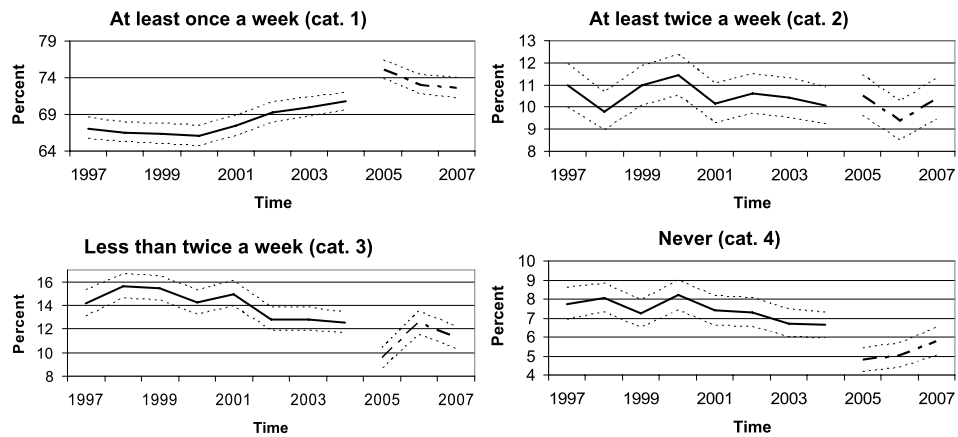}

\caption{Contact frequency with neighbors.
Solid line: observed series under the PSLC, dashed line: observed series
under the SSPEC, dotted line: 95\% confidence interval.}\label{fig2}
\end{figure}

The redesign from the PSLC to the SSPEC resulted in discontinuities in
most of the parameters about social participation and environmental
consciousness. As an example the series with the annual figures of the
parameters ``Separating chemical waste'' and ``Contact frequency with
neighbors'' are shown in Figures \ref{fig1} and \ref{fig2}, respectively. For both
variables it appears that there are significant discontinuities in two
or more of the underlying categories. The observed differences between
the last year of the PSLC in 2004 and the first year of the SSPEC in
2005 are summarized in Table \ref{tab1}. The observed differences between the
year before and the year after the changeover for other variables about
environmental consciousness and social participation are described in
the supplemental paper, van den Brakel and Roels (\citeyear{vandenBrakelRoels09}).

%f1 ###

%f2 ###

%
%t1 ###
\begin{table}[b]
\tabcolsep=0pt
\caption{Observed differences between the year before and the year after
the changeover for ``Separating chemical waste'' and ``Contact frequency
with neighbors''}\label{tab1}
\begin{tabular*}{\textwidth}{@{\extracolsep{\fill}}lccccc@{}}
\hline
% Row 1
& \multicolumn{5}{c}{\textbf{Category}}\\[-6pt]
& \multicolumn{5}{c}{\hrulefill}\\
% Row 2
\textbf{Variable}& \multicolumn{1}{c}{\textbf{1}} & \multicolumn{1}{c}{\textbf{2}} & \multicolumn
{1}{c}{\textbf{3}} & \multicolumn{1}{c}{\textbf{4}} & \multicolumn{1}{c}{\textbf{5}}\\
\hline
% Row 3
Freq. cont. neighb. & 4.38$^{\ast\ast}$ (0.90) & 0.46 (0.62)\phantom{.} & $-$2.99$^{\ast\ast}$ (0.63) & $-$1.84$^{\ast\ast}$ (0.47) &  \\
% Row 4
Sep. chemical waste & 2.26$^{\ast\ast}$ (0.89) & $-$5.25$^{\ast\ast}$ (0.50) & 0.79 (0.53)\phantom{.} & \hspace*{7pt}2.54$^{\ast\ast}$ (0.39) & $-$0.33 (0.54)\\
\hline
\end{tabular*}
\tabnotetext[]{}{$^{\ast}$: $p$-value $<0.05$; $^{\ast\ast}$: $p$-value $<0.01$. Standard errors in brackets.}
\end{table}

The observed differences are the results of the factors that changed
simultaneously in the survey redesign, real developments of the
parameter and sampling errors. The most important factors that changed
in the survey redesign are as follows:

\begin{itemize}
\item[$\bullet$] Differences between sampled target populations. The
SSPEC is based on a sample of persons aged 15 years and older with a
listed telephone number or cell-phone number. The PSLC is based on a
sample of persons aged 15 years and older. The SSPEC does not observe
the subpopulation without a listed telephone number or cell-phone
number. Additional analyses showed that this results in an
under-representation of young people and ethnic minorities. This
explains a substantial part of the discontinuities.

\item[$\bullet$] Differences in data collection modes. The SSPEC is a
telephone based survey, while in the PSLC data are collected in
face-to-face interviews conducted at the respondents' homes. Many
references in the literature emphasize that different collection modes
have systematic effects on the responses; see, for example, De Leeuw
(\citeyear{DeLeeuw05}) and Dillman and Christian (\citeyear{Dillman05}). These so-called model effects
arise for different reasons. Generally the interview speed in a
face-to-face interview is lower compared to an interview conducted by
telephone. Furthermore, respondents are more engaged with the interview
and are more likely to exert the required cognitive effort to answer
questions carefully in a face-to-face interview. Also, fewer socially
desirable answers are obtained under the CAPI mode due to the personal
contact with the interviewer. As a result, fewer measurement errors are
expected under the CAPI mode [Holbrook, Green and Krosnick (\citeyear{Holbrook03}) and Roberts
(\citeyear{Roberts07})].

\item[$\bullet$] Differences between data collection periods. The data
collection for the SSPEC is conducted in September through November,
while the PSLC is conducted continuously throughout the year. In the
series of the quarterly figures observed under the PSLC, seasonal
effects are observed in several parameters, which partially explain the
discontinuities.

\item[$\bullet$] Differences between questionnaire designs. Under the
PSLC, questions about social participation and environmental
consciousness were combined with questions about justice and crime
victimization in two different modules. Under the SSPEC, the questions
about social participation and environmental consciousness are
delineated in a new survey, which might have systematic effects on the
outcomes of these surveys [Kalton and Schuman (\citeyear{Kalton82}) and Dillman and
Christian (\citeyear{Dillman05})].

\item[$\bullet$] Differences between the contexts of the surveys. The
SSPEC is introduced as a survey that is focused on topics about social
participation and environmental consciousness. The PSLC is introduced as
a more general survey on living conditions. Subsequently, the survey
focuses on topics about justice, crime victimization, social
participation or environmental consciousness. This might have a
systematic selection effect on the respondents who decide to participate
in the survey. Furthermore, in the SSPEC the attention of the respondent
is completely focused on one topic, contrary to the PSLC, which also may
have systematic effects on the answer patterns of the respondents.

\end{itemize}

It is not immediately clear to what extent the differences summarized in
Table~\ref{tab1} are the result of a real change in the underlying phenomenon of
interest or are induced by the redesign of the survey. Even if no
significant difference is observed, it is still possible that a real
development could be nullified by an opposite redesign effect.

A general way to avoid confounding the autonomous development with
redesign effects is to conduct an experiment embedded in the ongoing
survey. If the effect of the separate factors that has changed in the
survey process should be quantified, then a factorial design should be
considered. Factorial designs or fractional factorial designs are
generally hard to combine with the fieldwork restrictions encountered in
the daily practice of survey sampling. Therefore, it is generally
necessary to combine the factors that changed in the redesign of the
survey into one treatment and test the total effect of all factors that
changed simultaneously in the redesign against the regular approach in a
two-treatment experiment. See van den Brakel (\citeyear{vandenBrakel08}) and van den Brakel,
Smith and Compton (\citeyear{vandenBrakelSmith08}) for a detailed discussion and alternative
approaches to quantify the effect of a survey redesign.

Since an experimental approach is not applied in this application, a
time series model is developed in the next section to quantify the total
effect of all factors that are modified in the survey redesign with the
purpose to avoid confounding with real developments of the respective
parameter. Some insight into the effect for some of the factors that
have changed in the survey redesign can be obtained by conducting
additional calculation on the existing data. The selection effect of
surveying the subpopulation that can be contacted by telephone can be
estimated with standard sampling theory for domain estimators from the
data obtained with the PSLC since this survey approaches the entire
population face-to-face. The effect of changing the period of data
collection can also be quantified by making, for example, quarterly
series for the PSLC and estimating the seasonal pattern. Due to the
relatively small sample sizes and the limited length of the series, it
turned out to be hard to establish significant seasonal effects.

%s4 ###
\section{Structural times series models} \label{sec:4}

In this section structural time series models are developed to estimate
the discontinuities in the series of a survey due to the redesign of the
underlying survey process. With a structural time series model, a series
is decomposed in a trend component, seasonal component, other cyclic
components, regression component and an irregular component. For each
component a stochastic model is assumed. This allows not only the trend,
seasonal and cyclic component but also the regression coefficients to be
time dependent. If necessary, ARMA components can be added to capture
the autocorrelation in the series beyond these structural components.
See Harvey (\citeyear{Harvey89}) or Durbin and Koopman~(\citeyear{Durbin01}) for details about
structural time series modeling.

%s4.1 ###
\subsection{Intervention analysis for time series obtained with
periodic surveys} \label{sec:4.1}

The variables of the PSLC and the SSPEC are defined as categorical
variables measured on an ordinal scale and the population values of
interest are the distributions in the population over the $K$ categories
of these variables. For each variable a $K$-dimensional vector
$\mathbf{y}_{t} = (y_{t,1},\ldots,y_{t,K})$ is defined where the elements of
$\mathbf{y}_{t}$ specify the proportions over the $K$ categories. Based
on the data observed under the PSLC and the SSPEC, direct estimates for
the unknown population values are obtained with the generalized
regression estimator. As a result, for each variable~$K$ series are
observed that specify the estimated proportions over $K$ categories and
are collected in the $K$-dimensional vector $\hat{\mathbf{y}}_{t} =
(\hat{y}_{t,1},\ldots,\hat{y}_{t,K}), t = 1, \ldots, T$.

Developing a time series model for survey estimates observed with a
periodic survey starts with a model, which states that the survey
estimate can be decomposed in the value of the population variable and a
sampling error: $\hat{y}_{t,k} = y_{t,k} + e_{t,k}$, with $e_{t,k}$ the
sampling error. Scott and Smith (\citeyear{Scott74}) proposed to consider the true
population value $y_{t,k}$ as the realization of a stochastic process
that can be properly described with a time series model. This approach
is applied to the series observed with the PSLC and the SSPEC using the
framework of structural time series modeling.

In classical sampling theory, it is generally assumed that the
observations obtained in the sample are true fixed values observed
without error; see, for example, Cochran (\citeyear{Cochran77}). This assumption is not
tenable if systematic differences are expected due to a redesign of the
survey process. van den Brakel and Renssen (\citeyear{vandenBrakel05}) proposed a measurement
error model for experiments embedded in sample surveys that link
systematic differences between a finite population variable observed
under different survey implementations. They consider the observed
population value obtained under a complete enumeration under two or more
different implementations of the survey process as the sum of a true
intrinsic value that is biased with a systematic effect induced by the
survey design, that is, $y_{t,k,l} = u_{t,k} + b_{k,l}$. Here $y_{t,k,l}$
is the population value of the $k$th parameter at time $t$ observed
under the $l$th survey approach, $u_{t,k}$ the true population value of
this parameter and $b_{k,l}$ the measurement bias induced by the $l$th
survey process used to measure $u_{t,k}$. The systematic difference
between two survey approaches is obtained by the contrast $y_{t,k,l} -
y_{t,k,l'} = b_{k,l} - b_{k,l'} \equiv\beta_{k}$. In the case of
embedded experiments, the systematic difference between two or more
survey approaches is estimated as the contrast between estimates
obtained from subsamples assigned to the different survey approaches. In
the time series approach, these differences are estimated using an
appropriate intervention variable. This allows for time dependent
differences. For notational convenience, the subscript $l$ will be
omitted in $y_{t,k,l}$, since the survey approach will be indicated
implicitly with the time period.

In the case of the PSLC and the SSPEC, a relatively short series for
annual data is considered. Therefore, the autonomous development of the
indicator that is described by the series is modeled with a stochastic
trend, a regression component and an irregular component. The regression
component consists of an intervention variable with a time independent
regression coefficient that describes the effect of the survey
transition. This approach is initially proposed by Harvey and Durbin~(\citeyear{Harvey00}). Seasonal, cyclic, ARMA and other auxiliary regression components
can be included in the model, for example, in the case of longer series
or monthly or quarterly data.

Based on the preceding considerations, the univariate structural time
series model for the $k$th component of $\hat{\mathbf{y}}_{t}$ is
defined as
%
%e1 ###
\begin{equation}\label{eq1}
\hat{y}_{t,k} = L_{t,k} + \beta_{k}\delta_{t} + \nu_{t,k} + e_{t,k}
\end{equation}
with $L_{t,k}$ a stochastic trend, $\delta_{t}$ an intervention variable
that describes under which survey the observations are obtained at
period $t$, $\beta_{k}$ the time independent regression coefficient for the
intervention variable, $\nu_{t,k}$ an irregular component for the time
series model of the population values $y_{t,k}$ and $e_{t,k}$ the
sampling error. It is assumed that the irregular component is normally
and independently distributed: $\nu_{t,k} \cong N(0,\sigma_{\nu} ^{2})$.

Surveys are often based on a rotating panel design. Such designs result
in partially overlapping samples with correlated sampling errors.
Particularly in these cases, a separate component for the sampling error
in the time series model might be required to capture this serial
correlation. Through this component the estimated variances for
the $\hat{y}_{t,k}$, which are generally available from the survey, can
be included in the time series model as prior information. Binder and
Dick (\citeyear{Binder90}) proposed the following general form for the sampling error
model to allow for nonhomogeneous variance in the sampling errors:
%
%e2 ###
\begin{equation}\label{eq2}
e_{t,k} = \omega_{t,k}\tilde{e}_{t,k},
\end{equation}
where $\omega_{t,k}$ is the standard error of $\hat{y}_{t,k}$ and
$\tilde{e}_{t,k}$ an ARMA process that models the serial correlation
between the sampling errors. Abraham and Vijayan (\citeyear{Abraham92}) and Harvey and
Chung (\citeyear{Harvey00}) applied MA models for the serial correlation in the
sampling errors. Pfeffermann (\citeyear{Pfeffermann91}),
Pfeffermann, Feder and Signorelli (\citeyear{Pfeffermann98})
and van den Brakel and Krieg (\citeyear{vandenBrakel09}) used AR models for the serial correlation in
the sampling errors. Autocorrelations can be estimated from the survey
data and can be used, like the design variances of $\hat{y}_{t,k}$, as
prior information in the sampling error model. Pfeffermann, Feder and Signorelli (\citeyear{Pfeffermann98})
developed a procedure to estimate the autocorrelation in the survey
errors from the separate panel estimates of a rotating panel design and
used this prior information to estimate the autocorrelation coefficients
of an AR model.

Generally there are systematic differences between the subsequent panels
of a rotating panel design. In the literature, this phenomenon is known
as rotation group bias (RGB) [Bailar (\citeyear{Bailar75})]. Pfeffermann (\citeyear{Pfeffermann91}) applied
a multivariate structural time series model to the series of the survey
estimates of the separate panel waves that accounts for this RGB and
applied an AR model for the autocorrelation of the sampling errors of
the different panels. Variances and autocorrelations of the sampling
errors are obtained by standard maximum likelihood estimation in this
application. van den Brakel and Krieg (\citeyear{vandenBrakel09}) used a multivariate
structural time series model similar to the model proposed by
Pfeffermann (\citeyear{Pfeffermann91}). They estimated the variances and autocorrelations of
the sampling errors from the survey data and used this as prior
information in the time series model.

The PSLC and the SSPEC are based on nonoverlapping cross-sectional
samples. The only difference between the sample designs is the yearly
sample size. As a result, there is no serial correlation between
sampling errors and nonhomogeneous variance is caused by differences in
the yearly sample size. Based on these considerations, it is decided to
combine both terms $\nu_{t,k}$ and $\varepsilon_{t,k}$ in one irregular
term, which is assumed to be normally and independently distributed with
zero mean and a variance that is inversely proportional to the sample
size:
%
%e3 ###
\begin{equation}\label{eq3}
\nu_{t,k} + e_{t,k} = \varepsilon_{t,k}, \qquad\varepsilon_{t,k} \cong N\biggl(0,\frac{\sigma_{\varepsilon
,k}^{2}}{n_{t}}\biggr).
\end{equation}
Defining the variance of the irregular term inversely proportional to
the sample size implies that it is implicitly assumed that the sampling
error dominates the irregular term. Note that the variance of
$\varepsilon_{t,k}$ is the variance of a binomial outcome and therefore
also depends on the value of $\hat{y}_{t,k}$. This could be taken into
account, for example, by taking $\operatorname{Var}(\varepsilon_{t,k}) =
\hat{y}_{t,k}(100 - \hat{y}_{t,k})/n_{t}$ or by including the estimated
standard error of $\hat{y}_{t,k}$ as prior information in the model
according to equation (\ref{eq2}). This aspect, however, is ignored in the
models used in this paper. It is also assumed that the irregular
components of (\ref{eq3}) at different time points are uncorrelated:
$\operatorname{Cov}(\varepsilon_{t,k}\varepsilon_{t',k}) = 0$ for $t \ne t'$. As a
result, model (\ref{eq1}) simplifies to
%
%e4 ###
\begin{equation}\label{eq4}
\hat{y}_{t,k} = L_{t,k} + \beta_{k}\delta_{t} + \varepsilon_{t,k}.
\end{equation}
For the stochastic trend, the widely applied smooth trend model is
assumed [see, e.g., Durbin and Koopman (\citeyear{Durbin01})]:
%
%
%e5 ###
\begin{eqnarray}
\label{eq5}
 L_{t,k} &=& L_{t - 1,k} + R_{t - 1,k},
 \nonumber
 \\[-8pt]
 \\[-8pt]
 \nonumber
R_{t,k} &=& R_{t
- 1,k} + \eta_{t,R,k},
\end{eqnarray}
with $L_{t,k}$ the level component and $R_{t,k}$ the stochastic slope
component of the trend and $\eta_{t,R,k}$ an irregular component. The
smooth trend model (\ref{eq5}) is a special case of the local linear trend
model, which also has an irregular term for $L_{t,k}$; see, for example,
Durbin and Koopman (\citeyear{Durbin01}), equation (3.2). The population values in
this application do not change rapidly over time. Therefore, a model
that gives smooth trend estimates seems to be appropriate. The choice
for (\ref{eq5}) also results in a more parsimonious model, which is an
additional advantage in this application where the length of the
observed series is small. It is assumed that the irregular components of
(\ref{eq5}) are normally and independently distributed, that is, $\eta_{t,R,k}
\cong N(0,\sigma_{R,k}^{2})$ and that they are uncorrelated at
different time points, that is, $\operatorname{Cov}(\eta_{t,R,k}\eta_{t',R,k}) =
0$ for $t \ne t'$. Furthermore, it is assumed that the irregular
components of (\ref{eq4}) and (\ref{eq5}) are uncorrelated:
$\operatorname{Cov}(\varepsilon_{t,k}\eta_{t',R,k}) = 0$ for all $t$ and $t'$.

The intervention variable models the effect of the survey redesign.
Three types of interventions are discussed: a level shift, a slope
intervention and an intervention on a seasonal pattern. Let $T_{R}$
denote the time period at which the survey process is redesigned. In the
case of a level intervention, it is assumed that the magnitude of the
discontinuity due to the survey redesign is constant over time. In this
case $\delta_{t}$ is defined as a dummy variable:
%
%e6 ###
\begin{equation}
\label{eq6}
\delta_{t} = \cases{
0, &\quad  $\mbox{if } t < T_{R},$ \cr
1, &\quad  $\mbox{if } t \ge T_{R}.$}
\end{equation}
In the case of a slope intervention, it is assumed that the magnitude of
the discontinuity increases over time. This is accomplished by defining
$\delta_{t}$ as
%
%e7 ###
\begin{equation}
\label{eq7}
\delta_{t} = \cases{
0, &\quad  $\mbox{if } t < T_{R},$ \cr
1 + t - T_{R}, &\quad  $\mbox{if } t \ge T_{R}.$}
\end{equation}
It is also possible to define an intervention on the seasonal or cyclic
pattern. Such interventions can be considered if an interaction is
expected between the survey redesign and the months or the quarters of
the year. In this case, a stochastic seasonal component is added to
equation (\ref{eq1}) or (\ref{eq4}). Widely applied models are trigonometric models and
the dummy variable seasonal model; see Durbin and Koopman~(\citeyear{Durbin01}),
Section 3.2, for expressions. Furthermore, the intervention variable
$\delta_{t}$ has the form (\ref{eq6}) and the regression coefficient $\beta_{k}$
is replaced by a time independent seasonal component.

The interventions described so far assume that the redesign only affects
the~point estimates of the survey. A survey redesign could, however,
also affect the variance of the measurement errors. An increase or
decrease of the variance of the measurement errors will be reflected in
the estimated variance of $\hat{y}_{t,k}$. A straightforward way to
account for such effects is to incorporate the estimated variances of
the survey estimates as prior information using sampling error model
(\ref{eq2}). Another possibility is to define separate model variances for the
irregular term $\varepsilon_{t,k}$ in the measurement equation for the
period before and after the implementation of the survey redesign, that
is, $\operatorname{Var}(\varepsilon_{t,k}) = \sigma_{\varepsilon,k,1}^{2}$ if $t <
T_{R}$ and $\operatorname{Var}(\varepsilon_{t,k}) = \sigma_{\varepsilon,k,2}^{2}$ if $t
\ge T_{R}$. The ratio between $\sigma_{\varepsilon,k,1}^{2}$ and
$\sigma_{\varepsilon,k,2}^{2}$ can be used to test hypotheses about the
equivalence of both variance components. This approach, however,
requires a sufficient number of observations under both surveys to test
the equivalence of these variance components with sufficient power.

The discontinuity in the series is modeled with an intervention variable
that describes the moment that the survey process is redesigned. This
approach assumes that the other components of the time series model
approximate the real development of the population variable reasonably
well and that there is no structural change in, for example, the trend
or the seasonal component at the moment that the new survey is
implemented. If a change in the real development of the population
variable exactly coincides with the implementation of the new survey,
then the model will wrongly assign this effect to the intervention
variable which is intended to describe the redesign effect. Information
available from series of correlated variables can be used to evaluate
the assumption that there is no structural change in the real evolution
of the population parameter. Such auxiliary series can also be added as
a regression component to the model, with the purpose to reduce the risk
that a structural change in the evolution of the series of the target
parameter is wrongly assigned to the intervention variable. An auxiliary
series can also be included as a dependent variable in a multivariate
model, which accounts for the correlation between the parameters of the
trend and seasonal components [Pfeffermann and Burck (\citeyear{Pfeffermann90}), Pfeffermann
and Bluer (\citeyear{Pfeffermann93})] or allows for a common trend [Harvey and Chung
(\citeyear{Harvey00})].

The risk that the intervention variable wrongfully absorbs a part of the
development of the real population value can be reduced by applying
parsimonious intervention parameters. Therefore, time dependent
interventions, like an intervention on a seasonal component, must be
applied carefully. These intervention parameters are more flexible and
will more easily absorb a part of the real evolution of the population
value, particularly if only a limited number of observations after the
survey changeover are available.

The intervention approach can be generalized in a straightforward way to
situations were the survey process has been redesigned at two or more
occasions. This is achieved by adding a separate intervention variable
for each time that the survey process has been modified.

%s4.2 ###
\subsection{State-space representation} \label{sec:4.2}

The structural time series models developed in Section \ref{sec:4.1} for the
separate parameters $\hat{y}_{t,k}$ of the vector
$\hat{\mathbf{y}}_{t}$ comprise a $K$-dimensional structural time series
model. The general way to proceed is to put this model in state-space
representation and analyze the model with the Kalman filter. The
state-space representation for this $K$-dimensional structural time
series model reads as
%
%e9 ###
%e8 ###
\begin{eqnarray}
\label{eq8}\hat{\mathbf{y}}_{t} &=& \mathbf{Z}_{t}\bolds{\alpha} _{t} +
\bolds{\varepsilon} _{t},\\
\label{eq9}\bolds{\alpha} _{t} &=& \mathbf{T}\bolds{\alpha} _{t - 1} + \bolds{\eta} _{t}.
\end{eqnarray}
The measurement equation (\ref{eq8}) describes how the observed series depends
on a vector of unobserved state variables $\bolds{\alpha}_{t}$ and a
vector with disturbances $\bolds{\varepsilon} _{t}$. The state vector
contains the level and slope components of the trend models and the
regression coefficients of the intervention variables. The transition
equation (\ref{eq9}) describes how these state variables evolve over time. The
vector $\bolds{\eta} _{t}$ contains the disturbances of the assumed
first-order Markov processes of the state variables. The matrices in (\ref{eq8})
and (\ref{eq9}) are given by
%
%e10 ###
\begin{subequation}\label{eq10}
\begin{eqnarray}
\label{eq10a}\bolds{\alpha} _{t} &=&
(L_{t,1},R_{t,1},\ldots,L_{t,K},R_{t,K},\beta_{1},\ldots,\beta_{K})^{T},\\
\label{eq10b}\mathbf{Z}_{t} &=& \bigl( \mathbf{I}_{[K]}
\otimes(1,0)|\delta_{t}\mathbf{I}_{[K]} \bigr), \\
\label{eq10c}\mathbf{T} &=& \operatorname{Blockdiag}\bigl(\mathbf{T}_{\mathrm{tr}},\mathbf{I}_{[K]}\bigr),
\\
\label{eq10d}\mathbf{T}_{\mathrm{tr}} &=& \mathbf{I}_{[K]} \otimes
\pmatrix{1 & 1 \cr 0
& 1 }
\end{eqnarray}
with $\mathbf{0}_{[p]}$ a column vector of order $p$ with each element
equal to zero and $\mathbf{I}_{[p]}$ the $p \times p$ identity matrix.
The disturbance vectors are defined as
\begin{eqnarray*}
\bolds{\varepsilon} _{t} &=&
(\varepsilon_{t,1},\ldots,\varepsilon_{t,K})^{T},
\\
\bolds{\eta} _{t} &=&
\bigl(0,\eta_{t,R,1},\ldots,0,\eta_{t,R,K},\mathbf{0}_{[K]}^{T}\bigr)^{T}.
\end{eqnarray*}
It is assumed that
\begin{eqnarray*}
E(\bolds{\varepsilon} _{t}) &=& \mathbf{0}_{[K]},\qquad
\operatorname{Cov}(\bolds{\varepsilon} _{t}) = \frac{1}{n_{t}}\operatorname{Diag}(\sigma_{\varepsilon
,1}^{2},\ldots,\sigma_{\varepsilon,K}^{2}),\\
E(\bolds{\eta} _{t}) &=& \mathbf{0}_{[3K]},\qquad\operatorname{Cov}(\bolds{\eta} _{t}) =
\operatorname{Diag}\bigl(0,\sigma_{R,1}^{2},\ldots,0,\sigma_{R,K}^{2},\mathbf{0}_{[K]}^{T}\bigr).
\end{eqnarray*}

In the case that each measurement equation and each transition equation
has its own separate hyperparameter, then (\hyperref[eq10]{10}) is a set of $K$
univariate structural time series models. If the measurement equations
or the transition equations share common hyperparameters, then (\hyperref[eq10]{10}) is a
$K$-dimensional seemingly unrelated multivariate structural time series
model. This is, for example, the case if $\sigma_{\varepsilon,1}^{2} =
\cdots = \sigma_{\varepsilon,K}^{2} = \sigma_{\varepsilon} ^{2}$.

The time independent regression coefficients of the intervention
variables are also included in the state vector, as described by Durbin
and Koopman~(\citeyear{Durbin01}), Section 6.2.2. The Kalman filter can be applied
straightforwardly to obtain estimates for the regression coefficients.
An alternative approach of estimating the regression coefficients is by
augmentation of the Kalman filter; see Durbin and Koopman (\citeyear{Durbin01}),
Section 6.2.3, for details.

In this application, each variable specifies the proportions over $K$
categories. In other words, each variable makes up a $K$-dimensional
series, which obeys the restriction that at each point in time these
series add up to one, that is, \mbox{$\sum_{k = 1}^{K} \hat{y}_{t,k} = 1$} and $0
\le\hat{y}_{t,k} \le1$. As a result, the $K$ regression coefficients
of the intervention variables must obey the restriction $\sum_{k = 1}^{K}
\beta_{k} = 0$. The multivariate structural time series model (\ref{eq10}) can
be augmented with this restriction by using the following design matrix
in the transition equation (\ref{eq9}):
\begin{equation}\label{eq10e}
\mathbf{T} = \operatorname{Blockdiag}(\mathbf{T}_{\mathrm{tr}},\mathbf{T}_{\mathrm{iv}}),
\end{equation}
where $\mathbf{T}_{\mathrm{tr}}$ is defined by (\ref{eq10d}) and
\begin{equation}\label{eq10f}
\mathbf{T}_{\mathrm{iv}} = \pmatrix{
\mathbf{I}_{[K - 1]} &
\mathbf{0}_{[K - 1]} \vspace*{4pt}\cr - \mathbf{1}_{[K - 1]}^{T} & 0}
\end{equation}
\end{subequation}
with $\mathbf{1}_{[p]}$ a column vector of order $p$ with each element
equal to one. Due to $\mathbf{T}_{\mathrm{iv}}$, defined in (\ref{eq10f}), the regression
coefficients as well as their Kalman-filter estimates obey the
restriction $\sum_{k = 1}^{K} \beta_{k} = 0$. In the case of a level
intervention, the time series after the moment of the survey transition
can be adjusted for the estimated discontinuities with $\tilde{y}_{t,k} =
\hat{y}_{t,k} - \hat{\beta} _{k}$. As an alternative, the series before
the survey transition can be adjusted with $\tilde{y}_{t,k} =
\hat{y}_{t,k} + \hat{\beta} _{k}$. In the case of a slope intervention,
the time series is adjusted with $\tilde{y}_{t,k} = \hat{y}_{t,k} -
\hat{\beta} _{k}\delta_{t}$. If the time series after the moment of the
survey transition is adjusted, then $\delta_{t}$ is defined by (\ref{eq7}). If
the time series before the changeover is adjusted, then $\delta_{t}$ is
defined as
%
%e11 ###
\begin{equation}
\label{eq11}
\delta_{t} = \cases{
t - T_{R}, & \quad $\mbox{if } t < T_{R},$ \cr
0, & \quad $\mbox{if }  t \ge T_{R}.$}
\end{equation}
Since the observed series and the estimated discontinuities obey the
required consistencies, the adjusted series does too.

An intervention on a seasonal component can be implemented in a way
similar to a level intervention. Let $s$ denote the number of time
periods of the seasonal set. The state vector $\bolds{\alpha}_{t}$ is
augmented with $K \times s$ state variables to model the seasonal
pattern for each parameter $\hat{y}_{t,k}$. The $K$ regression
coefficients $\beta_{k}$ are replaced by another set of $K \times
s$ state variables to model the intervention on seasonal pattern for each
target parameter. The design matrix of the measurement
equation~$\mathbf{Z}_{t}$ is augmented with a term $\mathbf{I}_{[K]}
\otimes\mathbf{z}_{[s]}^{T}$, where $\mathbf{z}_{[s]}$ is an
$s$-dimensional vector that describes the relation between the
observed series and the state variable of the trigonometric seasonal
model or the dummy variable seasonal model. Furthermore,
$\delta_{t}\mathbf{I}_{[K]}$ in $\mathbf{Z}_{t}$ is replaced
by $\delta_{t}\mathbf{I}_{[K]} \otimes\mathbf{z}_{[s]}^{T}$. The design
matrix of the transition equation is augmented with a block diagonal
element $\mathbf{I}_{[K]} \otimes\mathbf{T}_{s}$, where $\mathbf{T}_{s}$
denotes the transitional relation for a trigonometric model or the dummy
variable seasonal model. See Durbin and Koopman (\citeyear{Durbin01}), Section 3.2,
for expressions of~$\mathbf{z}_{[s]}$ and~$\mathbf{T}_{s}$. To force that
the sum over the seasonal intervention variables of the $K$ parameters
equals zero, the design matrix of the transition equation is augmented
with $\mathbf{T}_{\mathrm{iv}} \otimes\mathbf{T}_{s}$, where $\mathbf{T}_{\mathrm{iv}}$ is
defined by (\ref{eq10f}). Adjusted series are obtained with the approach
described for the level intervention.

%s4.3 ###
\subsection{Logratio transformations} \label{sec:4.3}

The multivariate model developed for $\hat{\mathbf{y}}_{t}$\break accounts
for
the restriction that $\sum_{k = 1}^{K} \hat{y}_{t,k} = 1$, but ignores
the restriction\break \mbox{$0 \le\hat{y}_{t,k} \le1$}. Ignoring the second
restriction might result in adjusted parameter estimates taking values
outside the admissible range $[0,1]$. In fact, each parameter defines a
set of time series that are observed on the ($K-1$)-dimensional
simplex. One way to account for both restrictions is to apply a logratio
transformation to the original data:
%
%e12 ###
\begin{equation}\label{eq12}
\hat{x}_{t,k} = \ln\biggl( \frac{\hat{y}_{t,k}}{\hat{y}_{t,K}} \biggr),\qquad  k=1, \ldots,
K-1.
\end{equation}
With (\ref{eq12}) the original observations $\hat{\mathbf{y}}_{t}$ are
transformed from the ($K-1$)-dimensional simplex to the
($K-1$)-dimensional real space; see Aitchison (\citeyear{Aitchison86}) for details.
State-space models are applied to logratio transformed compositional
time series obtained from repeated surveys by Silva and Smith (\citeyear{Silva01}).
They also give the details on how to account for serial correlation
between the sampling errors in logratio transformed survey data in the
case of partially overlapping surveys.

Instead of modeling the original series $\hat{\mathbf{y}}_{t}$ and
explicitly benchmarking the regression coefficients to restriction
(\ref{eq10f}), it is also possible to develop a set of $K-1$ univariate
structural time series models or a set of $K-1$ seemingly unrelated
structural time series for $\hat{\mathbf{x}}_{t} =
(\hat{x}_{t,1},\ldots,\hat{x}_{t,K - 1})^{t}$.

This model is obtained with formulas (\ref{eq8}) and (\ref{eq9}), where
$\hat{\mathbf{y}}_{t}$ is replaced by $\hat{\mathbf{x}}_{t}$ and taking
%
%e13 ###
\begin{eqnarray}\label{eq13}
\quad \bolds{\alpha} _{t} &=& (L_{t,1},R_{t,1},\ldots,L_{t,K - 1},R_{t,K -
1},\beta_{1},\ldots,\beta_{K - 1})^{T},\nonumber\\
\mathbf{Z}_{t} &=& \bigl( \mathbf{I}_{[K - 1]}
\otimes(1,0)|\delta_{t}\mathbf{I}_{[K - 1]} \bigr),\nonumber\\
\mathbf{T} &=&
\operatorname{Blockdiag}(\mathbf{T}_{\mathrm{tr}},\mathbf{T}_{\mathrm{iv}}),\qquad
\mathbf{T}_{\mathrm{tr}} = \mathbf{I}_{[K - 1]} \otimes
\pmatrix{ 1 & 1
\cr
0 & 1},\mathbf{T}_{\mathrm{iv}} = \mathbf{I}_{[K - 1]}, \\
\bolds{\varepsilon} _{t} &=& (\varepsilon_{t,1},\ldots,\varepsilon_{t,K -
1})^{T},\nonumber\\
\bolds{\eta} _{t} &=& \bigl(0,\eta_{t,R,1},\ldots,0,\eta_{t,R,K -
1},\mathbf{0}_{[K - 1]}^{T}\bigr)^{T}.\nonumber
\end{eqnarray}
The estimated discontinuities apply to the $K-1$ transformed series. In
the case of level intervention, the series observed after the survey
transition can be adjusted to the level of the series before the
changeover using $\tilde{x}_{t,k} = \hat{x}_{t,k} - \hat{\beta} _{k}$.
The series observed before the survey transition can be adjusted to the
level under the new situation with $\tilde{x}_{t,k} = \hat{x}_{t,k} +
\hat{\beta} _{k}$. In the case of a slope intervention, the time series
is adjusted with $\tilde{x}_{t,k} = \hat{x}_{t,k} - \hat{\beta}
_{k}\delta_{t}$. If the time series after the moment of the survey
transition is adjusted, then $\delta_{t}$ is defined by (\ref{eq7}). If the time
series before the changeover is adjusted, then $\delta_{t}$ is defined
by (\ref{eq11}). The state-space representation for a seasonal intervention
follows in a straightforward way from Section \ref{sec:4.2}. Subsequently, the
adjusted series can be transformed back to their original values that
specify the proportions over $K$ categories on the simplex by the
inverse of (\ref{eq12}), which is given by
%
%e14 ###
\begin{eqnarray}\label{eq14}
\tilde{y}_{t,k} &=& \frac{\exp(\tilde{x}_{t,k})}{\sum_{k = 1}^{K - 1}
\exp(\tilde{x}_{t,k}) + 1},\qquad k=1, \ldots, K-1,
\nonumber
\\[-8pt]
\\[-8pt]
\nonumber
\tilde{y}_{t,K} &=& \frac{1}{\sum_{k = 1}^{K - 1} \exp(\tilde{x}_{t,k})
+ 1}.
\end{eqnarray}
The adjusted series meets the consistency property that the adjusted
proportions add up to 1 and the values of the $K$ categories take values
in the range $[0,1]$, since the logratio transformation accounts for the
properties of the data observed on a simplex. The most important
drawback of this approach is that the interpretation of the results is
more difficult and the asymmetric treatment of the classes in the
logratio transformation (\ref{eq12}). Aitchison (\citeyear{Aitchison86}) shows that analysis
results obtained with logratio transformed compositional data are
invariant for the choice of the reference category that is used as the
denominator. This result is generalized to VARMA models applied to
logratio transformed compositional time series by Brunsdon and Smith
(\citeyear{Brunsdon98}) and state-space models by Silva and Smith (\citeyear{Silva01}). The outcomes
for the adjusted series, nevertheless, depend on the choice of the
category that is used in the denominator of the logratio transformation,
and can be attributed to the numerical optimization procedure used for
maximum likelihood estimation (see Section \ref{sec:4.5}).

The asymmetric treatment of the $K$ classes in logratio transformation
(\ref{eq12}) can be avoided by replacing the reference category $\hat{y}_{t,K}$
in the denominator by the geometric mean over the $K$ categories. This
results in the so-called central logratio transformation, which is
defined by
%
%e15 ###
\begin{equation}\label{eq15}
\hat{z}_{t,k} = \ln\biggl( \frac{\hat{y}_{t,k}}{g(\hat{y}_{t})} \biggr),\qquad k=1,\ldots,
K,
\end{equation}
with
%
%e16 ###
\begin{equation}\label{eq16}
g(\hat{y}_{t}) = \Biggl( \prod_{k = 1}^{K} \hat{y}_{t,k} \Biggr)^{{1}/{K}}.
\end{equation}
The advantage of this transformation is that the results do not depend
on the choice of a reference category. With (\ref{eq15}), however, the vector
$\hat{\mathbf{y}}_{t}$ is transformed from the ($K-1$)-dimensional
simplex to a linear subspace of the $K$-dimensional real space that is
confined by $\sum_{k = 1}^{K} \hat{z}_{t,k} = 0$.

The central logratio transformed series can be modeled with a
$K$-dimensional structural time series model. Since the $K$ regression\vspace*{1pt}
coefficients of the intervention variables must still obey the
restriction $\sum_{k = 1}^{K} \beta_{k} = 0$, time series model (\ref{eq8}),
(\ref{eq9}), (\ref{eq10a}) through (\ref{eq10f}) can be applied to model the series obtained
after the central logratio transformation. The series can be adjusted
for the estimated discontinuities in a similar way as described for the
untransformed and logratio transformed series. Subsequently, the
adjusted series can be transformed back to their original values by the
inverse of (\ref{eq15}):
%
%e17 ###
\begin{equation}\label{eq17}
\tilde{y}_{t,k} = \frac{\exp(\tilde{z}_{t,k})}{\sum_{k = 1}^{K} \exp
(\tilde{z}_{t,k})},\qquad k=1,\ldots, K.
\end{equation}

%s4.4 ###
\subsection{Benchmarking with series for subpopulations} \label{sec:4.4}

In sample surveys, parameter estimates for the total population are
often also itemized in different subpopulations or domains. The
following relationship applies between the series at the national level
and its breakdown in $H$ subpopulations:
%
%e18 ###
\begin{equation}\label{eq18}
\hat{y}_{t} = \sum_{h = 1}^{H} \frac{N_{h}}{N}\hat{y}_{t}^{h}.
\end{equation}
Here $\hat{y}_{t}^{h}$ and $N_{h}$ denote the parameter estimate and the
size of subpopulation $h$ respectively and $N = \sum_{h = 1}^{H} N_{h}$
the size of the total population. Applying the time series models,
described in Sections \ref{sec:4.1}, \ref{sec:4.2} and \ref{sec:4.3}, separately to the series at the
national level and its breakdown for these $H$ subpopulations might
result in inconsistencies between these series after adjustment for the
discontinuities. These inconsistencies arise since the regression
coefficients for the intervention variables do not account for the
consistency requirement specified by (\ref{eq18}).

One solution is to benchmark the adjusted series for the subpopulations
to the adjusted series at the national level, for example, by using a
Lagrange function. Let $\tilde{\mathbf{y}}_{t} =
(\tilde{\mathbf{y}}_{t,\mathrm{tot}}^{T},\tilde{\mathbf{y}}_{t,1}^{T},\ldots,\mathbf
{\tilde{y}}_{t,H}^{T})^{T}$
denote a $(H + 1)K$-vector containing the adjusted parameter estimates
for period $t$ for the total population $\tilde{\mathbf{y}}_{t,\mathrm{tot}} =
(\tilde{y}_{t,\mathrm{tot},1},\ldots,\tilde{y}_{t,\mathrm{tot},K})^{T}$ and the $H$
subpopulations $\tilde{\mathbf{y}}_{t,h} =
(\tilde{y}_{t,h,1},\ldots,\tilde{y}_{t,h,K})^{T}$. These parameters must
obey a set of linear restrictions such that (\ref{eq18}) is met and the unit sum
constraint for the vectors $\tilde{\mathbf{y}}_{t,\mathrm{tot}}$
and $\tilde{\mathbf{y}}_{t,h}$, for $h = 1, \ldots, H$, still applies.
This gives rise to a set of $(H + K)$ linear restrictions that can be
expressed as
%
%e19 ###
\begin{equation}\label{eq19}
\mathbf{R\tilde{y}}_{t}^{*} = \mathbf{c}
\end{equation}
with
%
%e21 ###
%e20 ###
\[
\mathbf{R} = \pmatrix{
\bigl(1, - \mathbf{f}_{[H]}^{T}\bigr)
\otimes\mathbf{L}\vspace*{2pt} \cr
\mathbf{I}_{[H + 1]} \otimes\mathbf{1}_{[K]}^{T}},\qquad
\mathbf{L} = \bigl(\matrix{
 \mathbf{I}_{[K - 1]} &
\mathbf{0}_{[K - 1]}}\bigr), \mathbf{f} = \biggl(
\frac{N_{1}}{N},\ldots,\frac{N_{H}}{N} \biggr)^{T}
\]
and
\[
\mathbf{c} = \bigl(
\mathbf{0}_{[K - 1]}^{T},\mathbf{1}_{[H + 1]}^{T} \bigr)^{T}.
\]
Applying the method of Lagrange multipliers gives
%
%e22 ###
\begin{equation}\label{eq20}
\tilde{\mathbf{y}}_{t}^{*} = \tilde{\mathbf{y}}_{t} +
\mathbf{VR}^{T}(\mathbf{RVR}^{T})^{ - 1}[\mathbf{c} -
\mathbf{R\tilde{y}}_{t}],
\end{equation}
where $\mathbf{V}$ denotes the covariance matrix
of $\tilde{\mathbf{y}}_{t}$. In (\ref{eq20}) the discrepancies $[\mathbf{c} -
\mathbf{R\tilde{y}}_{t}]$ are distributed over the values of
$\tilde{\mathbf{y}}_{t}$ proportional to their accuracy measure
specified by $\mathbf{V}$. This implies that the parameters for the total
population receive smaller adjustments than the parameters for the
subpopulations, since parameters for the total population are estimated
more precisely compared to domain estimates. The covariance matrix of
(\ref{eq20}) is given by
\[
\mathbf{V}(\tilde{\mathbf{y}}_{t}^{*}) = \mathbf{V} -
\mathbf{VR}^{T}(\mathbf{RVR}^{T})^{ - 1}\mathbf{RV}.
\]
The benchmarked estimates obtained with (\ref{eq20}) have smaller or equal
variances than the separately adjusted series. The interpretation of
this variance reduction is that the restrictions specified by (\ref{eq19}) add
additional information to the model that is applied to adjust the series
for the observed discontinuities.

Inconsistencies can also be avoided by modeling the untransformed series
for the total population and its breakdown in the $H$ subpopulations,
that is, $\hat{\mathbf{y}}_{t} =
(\hat{\mathbf{y}}_{t,\mathrm{tot}}^{T},\hat{\mathbf{y}}_{t,1}^{T},\ldots,\mathbf
{\hat{y}}_{t,H}^{T})^{T}$,
simultaneously in one multivariate model and including the consistency
requirements in the transition equation for the regression coefficient
of the intervention variables. To avoid unnecessary mathematical
notation, the transition equation is only given for the regression
coefficients of these intervention variables. The formulation of the
complete state-space representation follows directly from the models
defined in Section \ref{sec:4.1}.

Let $\bolds{\beta} = \mathbf{T}\bolds{\beta}$ denote the transition equation
for the time invariant regression coefficients of the intervention
variables for the series of the total population and the~$H$
subpopulations, that is, $\bolds{\beta} = ( \bolds{\beta}
_{\mathrm{tot}}^{T},\bolds{\beta} _{1}^{T},\ldots,\bolds{\beta} _{H}^{T} )^{T}$,
with $\bolds{\beta} _{\mathrm{tot}}$ the $K$-dimensional vector containing the
intervention variables for the $K$ categories of the parameter for the
total population and $\bolds{\beta} _{h}$ the $K$-dimensional vector
containing the intervention variables of the parameter for the $h$th
subpopulation. If the transition matrix is defined as
\[
\mathbf{T} = \pmatrix{
\mathbf{O}_{[K \times K]} &
\mathbf{f}_{[H]}^{T} \otimes\mathbf{T}_{\mathrm{iv}} \vspace*{3pt}\cr \mathbf{1}_{[H]}
\otimes\mathbf{O}_{[K \times K]} & \mathbf{I}_{[H]}
\otimes\mathbf{T}_{\mathrm{iv}}},
\]
where $\mathbf{T}_{\mathrm{iv}}$ is defined by (\ref{eq10f}), then it follows that the
adjusted series meet the consistencies specified by (\ref{eq18}) as well as the
unit sum constraint for the $K$ classes of the parameter for the total
population and the $H$ subpopulations.

Both methods can be generalized to benchmark the series for the
population total and two or more domain classifications simultaneously.
Adding too many restrictions, however, might result in numerical
problems for solving (\ref{eq20}) or estimating the state-space model.

%s4.5 ###
\subsection{Implementation of the Kalman filter} \label{sec:4.5}

After having expressed the multivariate structural time series model in
state-space representation and under the assumption of normally
distributed error terms, the Kalman filter can be applied to obtain
optimal estimates for the state variables as well as the measurement
equation; see, for example, Durbin and Koopman (\citeyear{Durbin01}). Estimates for
state variables for period $t$ based on the information available up to
and including period $t$ are referred to as the filtered estimates. The
filtered estimates of past state vectors can be updated if new data
become available. This procedure is referred to as smoothing and results
in smoothed estimates that are based on the completely observed time
series. So the smoothed estimate for the state vector for period $t$
also accounts for the information made available after time period $t$.
In this paper, point estimates and standard errors for the state
variables are based on the smoothed Kalman-filter estimates using the
fixed interval smoother. See Harvey (\citeyear{Harvey89}) or Durbin and Koopman (\citeyear{Durbin01})
for technical details.

The nonstationary state variables are initialized with a diffuse prior,
that is, the expectations of the initial states are equal to zero and
the initial covariance matrix of the states is diagonal with large
diagonal elements. The time independent regression coefficients of the
intervention variables are also initialized with a diffuse prior, as
described by Durbin and Koopman (\citeyear{Durbin01}), Section 6.2.2.

The analysis is conducted with software developed in Ox in combination
with the subroutines of SsfPack 3.0; see Doornik (\citeyear{Doornik98})
and Koopman, Shephard and Doornik (\citeyear{Koopman99,Koopman08}). In SsfPack 3.0 an exact diffuse log-likelihood
function is obtained with the procedure proposed by Koopman (\citeyear{Koopman97}).
Maximum likelihood estimates for the hyperparameters, that is, the
variance components of the stochastic processes for the state variables,
are obtained using a numerical optimization procedure [BFGS algorithm,
Doornik (\citeyear{Doornik98})]. To avoid negative variance estimates, the
log-transformed variances are estimated. The Ox-program, used to conduct
the analyses, is available as a supplemental file, van den Brakel and
Roels~(\citeyear{vandenBrakelRoels09}).

%s5 ###
\section{Results} \label{sec:5}

%s5.1 ###
\subsection{Results with four different time series models} \label{sec:5.1}

The time series models developed in Section \ref{sec:4} are applied to the series
of ``Separating chemical waste'' and ``Contact frequency with
neighbors,'' which are plotted in Figures \ref{fig1} and \ref{fig2}. The results obtained
with four different models are compared. These models assume that the
series can be decomposed in a stochastic trend, a level intervention and
an irregular term. Because the series concern annual data, it was not
necessary to use a seasonal component. This allowed the selection of
very parsimonious models, which was inevitable since the series are very
short (11 years). Adding AR or MA components deteriorated the model fits
and generally resulted in overfitting of the data.

The first model, denoted M1, is a seemingly unrelated structural time
series model applied to the untransformed series. This model is defined
by equations (\ref{eq6}), (\ref{eq8}), (\ref{eq9}), (\ref{eq10a}), (\ref{eq10b}), (\ref{eq10c}) and (\ref{eq10d}). Note that
there is no restriction for the estimated discontinuities. This is a
seemingly unrelated structural time series model, since it is assumed
that the variances of the irregular terms in the measurement equations
are equal, that is, $\sigma_{\varepsilon,1}^{2} = \cdots =
\sigma_{\varepsilon,K}^{2} = \sigma_{\varepsilon} ^{2}$. Due to the
limited length of the series, this assumption is made to reduce the
number of hyperparameters to be estimated.

The second model, denoted M2, is the restricted multivariate model
defined by equations (\ref{eq6}), (\ref{eq8}), (\ref{eq9}), (\ref{eq10a}), (\ref{eq10b}), (\ref{eq10d}), (\ref{eq10e}) and
(\ref{eq10f}). The observed series are not transformed and the regression
coefficients of the intervention variables are explicitly benchmarked by
restriction $\mathbf{T}_{\mathrm{iv}}$ defined in (\ref{eq10f}). It is also assumed
that $\sigma_{\varepsilon,1}^{2} = \cdots = \sigma_{\varepsilon,K}^{2} =
\sigma_{\varepsilon} ^{2}$.

The third model, denoted M3, is a seemingly unrelated structural time
series model applied to the $K-1$ series obtained after applying
logratio transformation~(\ref{eq12}) using the last category as the reference
category in the denominator. This model is defined by (\ref{eq6}), (\ref{eq8}), (\ref{eq9}) and
(\ref{eq13}). To reduce the number of hyperparameters, it is assumed that
$\sigma_{\varepsilon,1}^{2} = \cdots = \sigma_{\varepsilon,K - 1}^{2} =
\sigma_{\varepsilon} ^{2}$.

The fourth model, denoted M4, is the restricted multivariate model
applied to the $K$ series obtained after applying the central logratio
transformation (\ref{eq15}). This model is defined by equations (\ref{eq6}), (\ref{eq8}), (\ref{eq9}),
(\ref{eq10a}), (\ref{eq10b}), (\ref{eq10d}), (\ref{eq10e}) and (\ref{eq10f}). It is assumed
that $\sigma_{\varepsilon,1}^{2} = \cdots = \sigma_{\varepsilon,K - 1}^{2}
= \sigma_{\varepsilon} ^{2}$.

For each model two analyses are conducted. One is based on the data
available up to and including 2006, the other on the complete series,
including 2007. This gives some intuition of the size of the revision of
the estimate of the discontinuity if an additional observation under the
new approach becomes available.

Estimation results for the discontinuities under the different models
are given in Table \ref{tab2} for the parameter ``Separating chemical waste'' and
in Table \ref{tab3} for the parameter ``Contact frequency with neighbors.''

As expected in advance, the estimated discontinuities under M1 do not
obey the restriction $\sum_{k = 1}^{K} \hat{\beta} _{k} = 0$. As a
result, the corrected series are not consistent, since the categories
for a parameter do not add up to one.

The multivariate model for the original series (M2) and the central
logratio transformed series (M4) results in consistent series since the
estimates for the discontinuities are forced to obey the required
restriction. Augmenting the model with restriction (\ref{eq10f}) also reduces
the standard errors of the estimated discontinuities, since the
restriction adds additional information to the model. This follows if
the results obtained with the multivariate model (M2) are compared with
the results obtained with the seemingly unrelated time series model (M1)
for the original series.

Another way to preserve the consistency between the series of the $K$
categories of a parameter is to apply the logratio transformation, since
this transformation eliminates the redundancy due to the unit sum
constraint over the $K$ categories. The estimated discontinuities for
the logratio and central logratio transformation in Tables \ref{tab2} and \ref{tab3} are
the results obtained with the transformed series.

%
%t2 ###
\begin{table}
\caption{Estimated discontinuities for ``Separating chemical waste''
with different models}
\label{tab2}
\begin{tabular*}{\textwidth}{@{\extracolsep{4in minus 4in}}lcd{2.8}d{2.8}d{1.8}d{2.8}c@{}}
\hline
% Row 1
 & & \multicolumn{5}{c@{}}{\textbf{Category}}\\[-6pt]
  & & \multicolumn{5}{c@{}}{\hrulefill}\\
% Row 2
\textbf{Model}& $\bolds{T}$& \multicolumn{1}{c}{\textbf{1}} & \multicolumn{1}{c}{\textbf{2}} &
\multicolumn{1}{c}{\textbf{3}} & \multicolumn{1}{c}{\textbf{4}} & \multicolumn{1}{c@{}}{\textbf{5}}\\
\hline
% Row 3
M1 & 2006 & 4.29\mbox{ }(1.21) & -4.34\mbox{ }(1.21) & 0.00\mbox{ }(1.21) & 1.50\mbox{ }(1.21) & $-$1.44\mbox{ }(1.21)\\
% Row 4
M1 & 2007 & 1.91\mbox{ }(1.88) & -4.15\mbox{ }(0.77) & -0.07\mbox{ }(0.77) & 1.49\mbox{ }(0.77) & $-$1.17\mbox{ }(0.98)\\
% Row 5
M2 & 2006 & 4.29\mbox{ }(1.07) & -4.35\mbox{ }(1.07) & -0.01\mbox{ }(1.07) & 1.50\mbox{ }(1.07) & $-$1.44\mbox{ }(1.07)\\
% Row 6
M2 & 2007 & 3.07\mbox{ }(1.44) & -4.01\mbox{ }(0.75) & 0.07\mbox{ }(0.75) & 1.63\mbox{ }(0.75) & $-$0.76\mbox{ }(0.98)\\
% Row 7
M3$^\ast$ & 2006 & -0.06\mbox{ }(0.14) & -1.08\mbox{ }(0.20) & 0.16\mbox{ }(0.10) & 1.00\mbox{ }(0.20) & \\
% Row 8
M3$^\ast$ & 2007 & 0.19\mbox{ }(0.15) & -0.77\mbox{ }(0.21) & 0.23\mbox{ }(0.11) & 0.68\mbox{ }(0.12) & \\
% Row 9
M4$^\ast$ & 2006 & -0.04\mbox{ }(0.26) & -1.06\mbox{ }(0.26) & 0.22\mbox{ }(0.31) & 1.01\mbox{ }(0.16) & $-$0.13\mbox{ }(0.07)\\
% Row 10
M4$^\ast$ & 2007 & -0.05\mbox{ }(0.25) & -1.09\mbox{ }(0.26) & 0.17\mbox{ }(0.30) & 1.00\mbox{ }(0.21) & $-$0.03\mbox{ }(0.07)\\
\hline
\end{tabular*}
\tabnotetext[]{}{$^\ast$: Results obtained for the (central) logratio transformed series. $T$:
Period of the last observation included in the analysis. Standard errors
in brackets.}
\end{table}
%
%t3 ###
\begin{table}[b]
\caption{Estimated discontinuities for ``Contact frequency neighbors''
with different models}
\label{tab3}
\begin{tabular*}{\textwidth}{@{\extracolsep{\fill}}lcd{1.8}d{2.8}d{2.8}c@{}}
\hline
% Row 1
 & &\multicolumn{4}{c@{}}{\textbf{Category}}\\[-6pt]
  & &\multicolumn{4}{c@{}}{\hrulefill}\\
% Row 2
\textbf{Model} & $\bolds{T}$ & \multicolumn{1}{c}{\textbf{1}} & \multicolumn{1}{c}{\textbf{2}} & \multicolumn
{1}{c}{\textbf{3}} & \multicolumn{1}{c@{}}{\textbf{4}}\\
\hline
% Row 3
M1 & 2006 & 4.79\mbox{ }(1.19) & 0.31\mbox{ }(0.69) & -4.19\mbox{ }(1.32) & \phantom{0.}1.60\mbox{ }(0.51)\\
% Row 4
M1 & 2007 & 4.40\mbox{ }(1.20) & -0.09\mbox{ }(0.59) & -3.18\mbox{ }(1.30) & $-$1.36\mbox{ }(0.59)\\
% Row 5
M2 & 2006 & 5.02\mbox{ }(0.93) & 0.46\mbox{ }(0.66) & -3.92\mbox{ }(0.96) & $-$1.56\mbox{ }(0.48)\\
% Row 6
M2 & 2007 & 4.44\mbox{ }(0.93) & -0.07\mbox{ }(0.56) & -3.01\mbox{ }(0.95) & $-$1.35\mbox{ }(0.56)\\
% Row 7
M3$^\ast$ & 2006 & 0.33\mbox{ }(0.09) & 0.27\mbox{ }(0.09) & 0.16\mbox{ }(0.09) &  \\
% Row 8
M3$^\ast$ & 2007 & 0.38\mbox{ }(0.11) & 0.30\mbox{ }(0.10) & 0.14\mbox{ }(0.08) & \\
% Row 9
M4$^\ast$ & 2006 & 0.14\mbox{ }(0.06) & 0.08\mbox{ }(0.06) & -0.03\mbox{ }(0.06) & $-$0.19\mbox{ }(0.06)\\
% Row 10
M4$^\ast$ & 2007 & 0.12\mbox{ }(0.05) & 0.07\mbox{ }(0.05) & -0.03\mbox{ }(0.05) & $-$0.16\mbox{ }(0.05)\\
\hline
\end{tabular*}
\tabnotetext[]{}{$^\ast$: Results obtained for the (central) logratio transformed series. $T$:
Period of the last observation included in the analysis. Standard errors
in brackets.}
\end{table}

The results obtained under equivalent models illustrate the size of the
revision for the estimated discontinuities if the data for an additional
year becomes available. Adding the estimates obtained in 2007 to the
series results in a revision of the estimated discontinuities. Large
revisions are observed for the first category of ``Separating chemical
waste'' under model M1 and the fourth category of ``Contact frequency
with neighbors'' under model M1. For the other three models the sizes of
the revisions are smaller with respect to the standard errors. It can be
expected that the size of the revisions decreases if the length of the
series increases, particularly if the number of data points after the
changeover increases.

\begin{figure}

\includegraphics{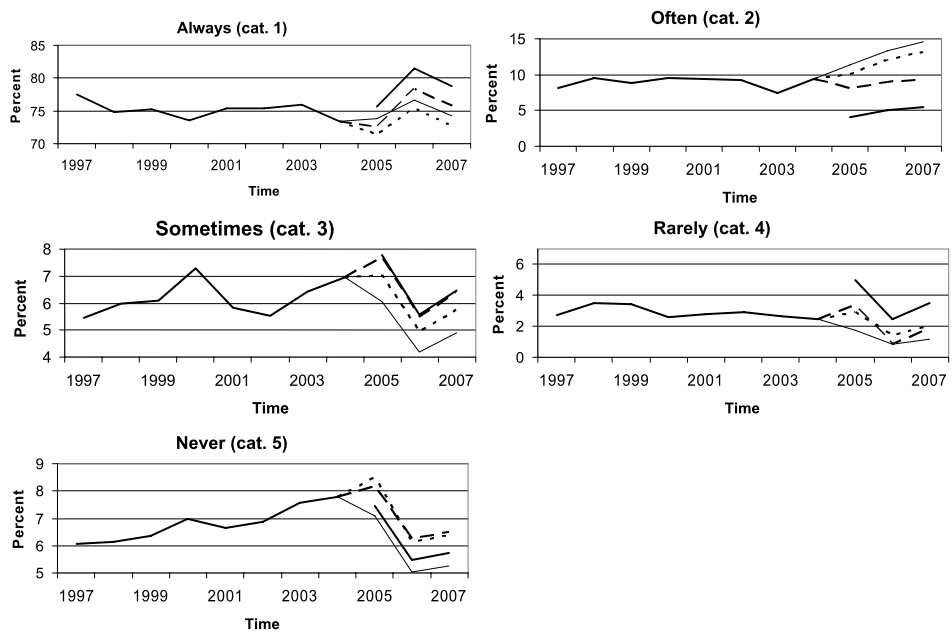}

\caption{Separating chemical waste.
Solid line 1997--2004 estimate based on the PSLC, solid line 2005--2007
estimate based on the SSPEC, dotted line corrected series based on a
logratio transformation, dashed line corrected series based on
untransformed data, thin solid line corrected series based on central
logratio transformation.}\label{fig3}
\end{figure}

\begin{figure}

\includegraphics{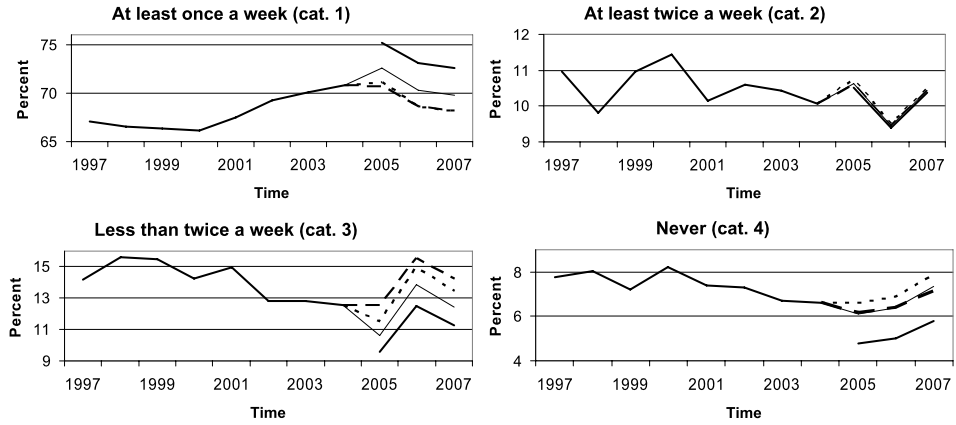}

  \caption{Contact frequency with neighbors.
Solid line 1997--2004 estimate based on the PSLC, solid line 2005--2007
estimate based on the SSPEC, dotted line corrected series based on a
logratio transformation, dashed line corrected series based on
untransformed data, thin solid line corrected series based on central
logratio transformation.}\label{fig4}
\end{figure}

The original data, the corrected series obtained with models M2, M3 and
M4, are shown in Figures \ref{fig3} and \ref{fig4}. The outcomes obtained under the SSPEC
for the period 2005 through 2007 are corrected to make the series
comparable with the outcomes of the PSLC, using the procedure described
in Section \ref{sec:4}. In Section \ref{sec:5.2} a simulation study is conducted to
investigate which model is most appropriate to estimate discontinuities
and produce corrected series for the variables of the PSLC and the
SSPEC.

%f3 ###

%f4 ###

%s5.2 ###
\subsection{Model evaluation} \label{sec:5.2}

The underlying assumptions of the state-space model are that the
disturbances of the measurement and system equations are normally
distributed and serially independent with constant variances. There are
different diagnostic tests available in the literature to test to what
extent these assumptions are met; see Durbin and Koopman (\citeyear{Durbin01}), Section
2.12.

In this application model evaluation is particularly important. The
observed series are the outcome of variables that have a multinomial
response at each time period. The Gaussian models M1 and M2 are applied
to the untransformed data and therefore do not account for this
property. Models M3 and M4 are also Gaussian, but account for the
multinomial response through the logratio and a central logratio
transformation. Durbin and Koopman (\citeyear{Durbin00}) and Durbin and Koopman (\citeyear{Durbin01}),
Chapters 10 and 11, describe simulation methods for the analysis of
non-Gaussian models and can be used as an alternative.

Another point of concern is the limited length of the available series.
Only 11 periods are observed, which might affect the precision of the
maximum likelihood estimates for the hyperparameters and the smoothed
Kalman-filter estimates for the discontinuities. Furthermore, standard
diagnostic tests to evaluate model assumptions will not have sufficient
power to asses model deficiencies and are therefore not very useful in
this application. As an alternative, two simulations are conducted.

%s5.2.1 ###
\subsubsection{Simulation with different time series lengths} \label{sec:5.2.1}

In the first simulation the effect of the length of the series on the
reliability of the estimates for the hyperparameters and the
discontinuities is investigated. Replications of time series are
generated from the unconditional distribution implied by model M3 using
the maximum likelihood estimates for the hyperparameters and the
smoothed estimates for the discontinuities obtained for the variable
``Contact frequency with neighbors.''

For each replication, states and observations are generated using the
SsfPack procedure SsfRecursion as described in Koopman, Shephard and Doornik (\citeyear{Koopman08}),
Section \ref{sec:4.1}. This procedure uses standard normal random numbers for the
disturbance terms of the measurement and system equations. The maximum
likelihood estimates for the hyperparameters and the smoothed estimates
for the discontinuities are used to define the state-space model.
Subsequently, model M3 is applied to analyze the simulated time series.

Three different simulations are conducted. In the first simulation, time
series with a length of 11 observations, 8 before and 3 after the survey
redesign, are generated. In the second simulation, time series with a
length of 22 observations, 16 before and 6 after the survey redesign,
are generated. In the third simulation, time series with a length of 44
observations, 32 before and 12 after the survey redesign, are generated.
The variance of the irregular terms of the measurement equation is
inversely proportional to the yearly sample size of the survey. For the
first simulation the actual sample sizes of the PSLC and the SSPEC are
used. In the second and the third simulation additional sample sizes are
generated from a uniform distribution where the minimum and maximum
yearly sample size of the PSLC and the SSPEC are used as the lower and
upper boundaries of the uniform distribution. For each simulation study
10,000 time series are generated.

\begin{figure}

\includegraphics{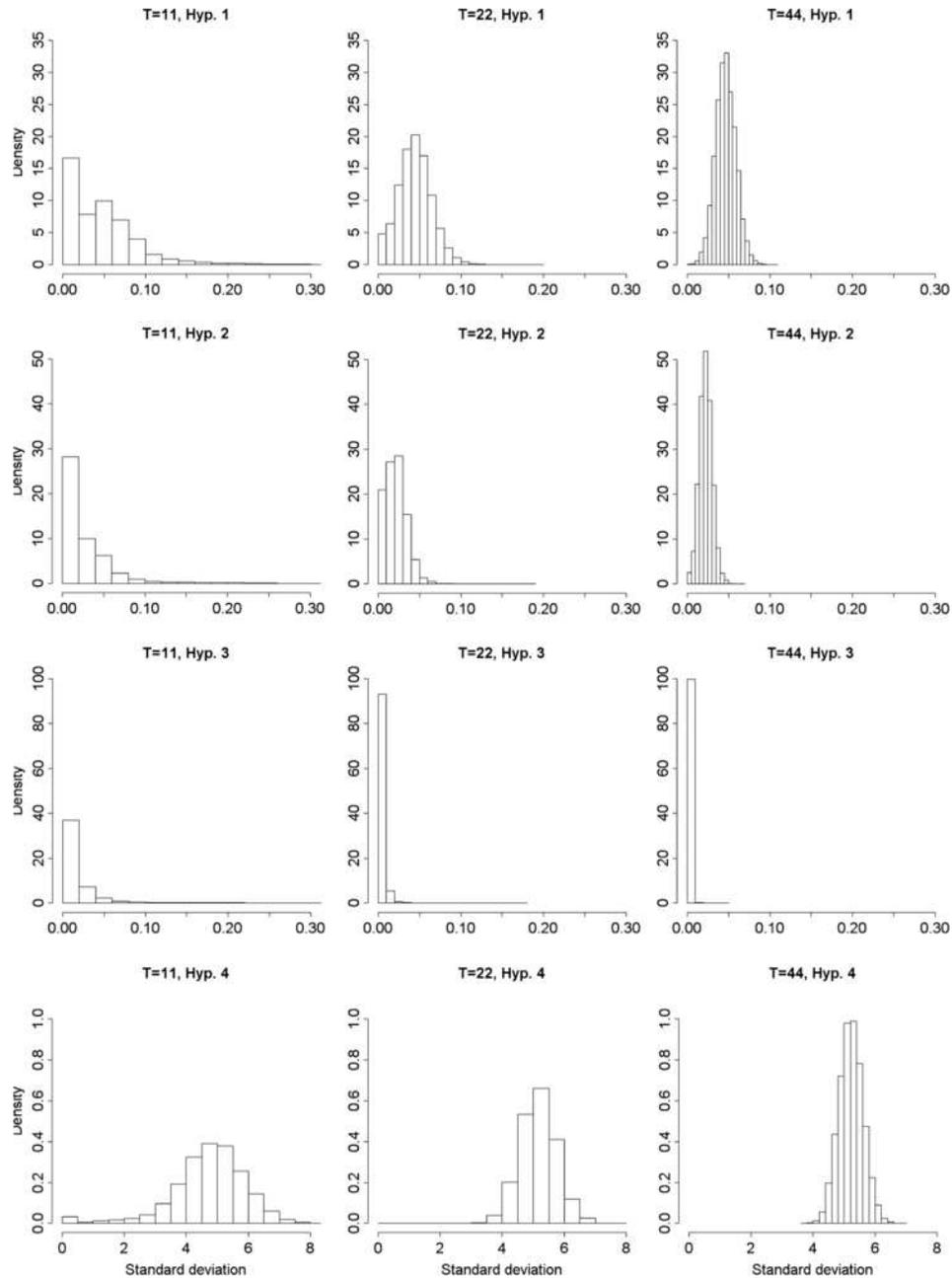}

  \caption{Resample distributions estimated hyperparameters for different
time series lengths.
Hyp.~1, Hyp. 2, Hyp. 3: Standard deviations irregular terms of the slope
from the trend model for three series obtained after logratio
transformation, that is, $\sigma_{R,1}, \sigma_{R,2},\sigma_{R,3}$.
Hyp. 4: Standard deviation irregular terms of the measurement equations,
that is, $\sigma_{\varepsilon}$.}\label{fig5}
\end{figure}

\begin{figure}

\includegraphics{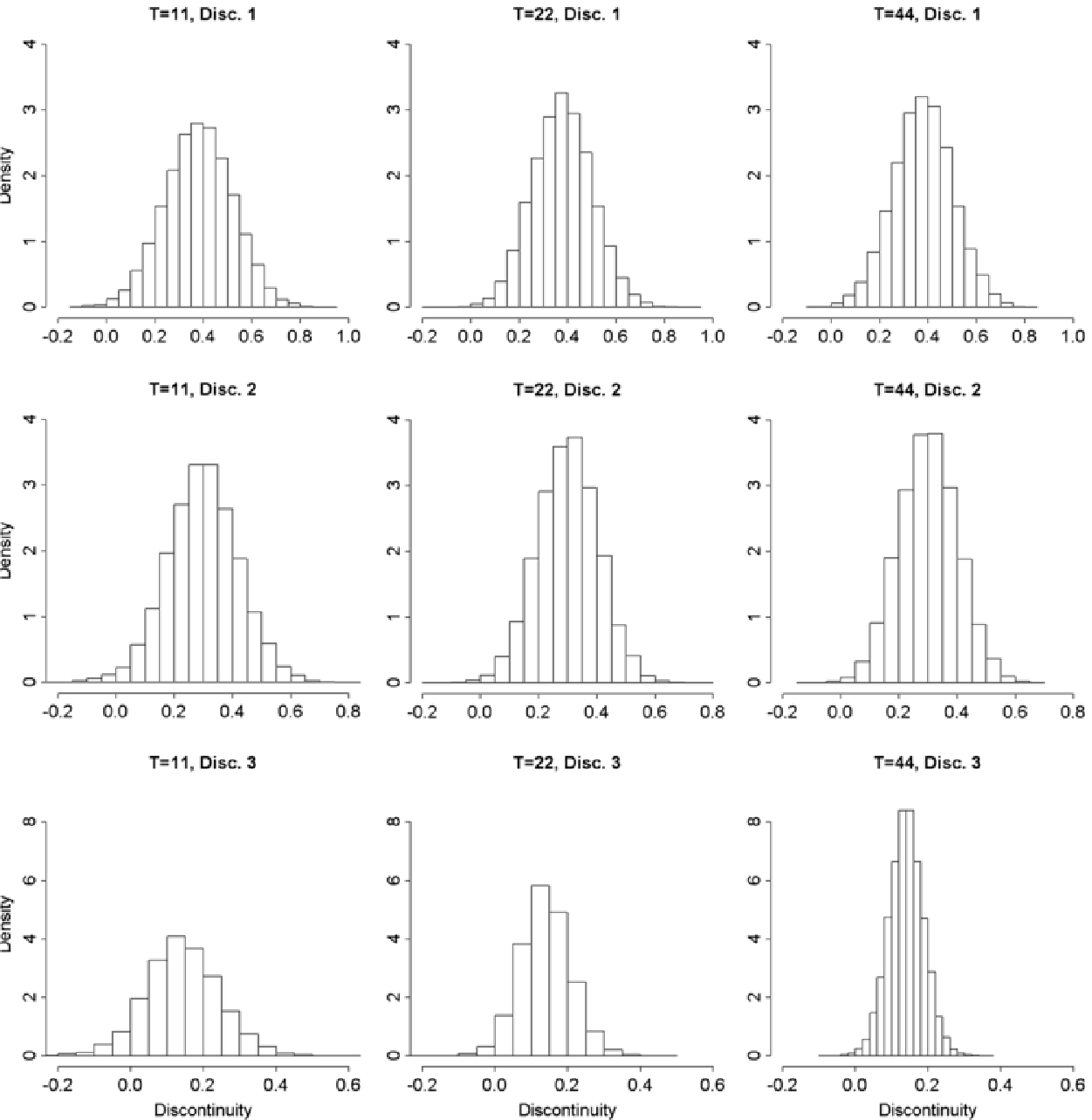}

  \caption{Resample distributions estimated discontinuities for different
time series lengths.
Disc. 1, Disc. 2, Disc. 3: Discontinuity for three series obtained after
logratio transformation, that is, $\beta_{1}$, $\beta_{2}$, $\beta_{3}$.}\label{fig6}
\end{figure}

The resample distributions of the maximum likelihood estimates for the
hyperparameters and the smoothed estimates for the discontinuities are
used to obtain more insight in the reliability of these model estimates
in this application where only a limited number of data points are
available. In Table \ref{tab4} the means and standard errors of the resample
distributions of the estimated hyperparameters and discontinuities are
compared with the values used in the assumed distribution. Standard
errors are obtained with the resample standard deviation. The resample
distributions of the estimated hyperparameters and discontinuities are
plotted in Figures \ref{fig5} and \ref{fig6}.

%
%t4 ###
\begin{table}[b]
\caption{Simulation results for the estimated hyperparameters and
discontinuities with different lengths of the times series}
\label{tab4}
\begin{tabular*}{\textwidth}{@{\extracolsep{\fill}}ld{1.4}d{1.12}d{1.12}c@{}}
\hline
% Row 1
& & \multicolumn{3}{c@{}}{\textbf{Simulated values}}\\[-6pt]
& & \multicolumn{3}{c@{}}{\hrulefill}\\
% Row 2
\textbf{Parameter} &  \multicolumn{1}{c}{\textbf{Real values}} & \multicolumn{1}{c}{$\bolds{T=11}$} & \multicolumn
{1}{c}{$\bolds{T=22}$} & \multicolumn{1}{c}{$\bolds{T=44}$}\\
\hline
% Row 3
Hyp. 1 & 0.0480 & 0.0460\mbox{ }(0.0464) & 0.0445\mbox{ }(0.0208) & 0.0467\mbox{ }(0.0123)\\
% Row 4
Hyp. 2 & 0.0237 & 0.0261\mbox{ }(0.0412) & 0.0210\mbox{ }(0.0139) & 0.0227\mbox{ }(0.0079)\\
% Row 5
Hyp. 3 & 0.000 & 0.0170\mbox{ }(0.0392) & 0.0027\mbox{ }(0.0064) & 0.0006\mbox{ }(0.0014)\\
% Row 6
Hyp. 4 & 5.260 & 4.7182\mbox{ }(1.2177) & 5.1664\mbox{ }(0.5833) & 5.2223\mbox{ }(0.3869)\\
% Row 7
Disc. 1 & 0.380 & 0.380\mbox{ }(0.141) & 0.378\mbox{ }(0.124) & 0.379\mbox{ }(0.123)\phantom{00}\\
% Row 8
Disc. 2 & 0.300 & 0.298\mbox{ }(0.122) & 0.300\mbox{ }(0.105) & 0.300\mbox{ }(0.101)\phantom{00}\\
% Row 9
Disc. 3 & 0.140 & 0.142\mbox{ }(0.104) & 0.139\mbox{ }(0.070) & 0.140\mbox{ }(0.049)\phantom{00}\\
\hline
\end{tabular*}
\tabnotetext[]{}{Hyp. 1, Hyp. 2, Hyp. 3: Standard deviations irregular terms of the slope
from the trend model for three series obtained after logratio
transformation, that is, $\sigma_{R,1}, \sigma_{R,2},\sigma_{R,3}$. Hyp.
4: Standard deviation irregular terms of the measurement equations, that
is, $\sigma_{\varepsilon}$. Disc. 1, Disc. 2, Disc. 3: Discontinuity
for three series obtained after logratio transformation, that is,
$\beta_{1}, \beta_{2},\beta_{3}$. Standard errors in brackets.}
\end{table}

%f5 ###

%f6 ###

The absolute difference between the real value and the mean of the
resample estimates for the hyperparameters and the discontinuities can
be considered as a measure for unbiasedness. The standard error of the
mean of the resample estimates can be taken as a measure for the
precision. The differences between the real value and the mean of the
resample estimates are small with respect to the standard error for
different lengths of the time series. This implies that there are no
indications that a limited number of observations results in biased
parameter estimates. The precision of the maximum likelihood estimates
of the hyperparameters clearly improves with the length of the time
series. It follows from Table \ref{tab4} that the size of the standard errors
decreases with the length of the series. The same conclusion follows
from Figure \ref{fig5}. Short series result in wide and skewed resample
distributions around the true values. The resample distributions center
on the true value and become more symmetrical if the length of the
series increases. The precision of the smoothed estimates of the
discontinuities, on the other hand, is much better in the case of the
shortest time series. It can be seen from Table \ref{tab4} that the decrease of
the standard errors if the length of the series increases is much
smaller compared to the hyperparameters. The same conclusion follows
from Figure \ref{fig6}. The effect of the length of the series on the dispersion
of the resample distribution around the true values is much smaller. The
sample distributions are also allocated more symmetrically around the
true values, even in the case of the shortest time series.

%s5.2.2 ###
\subsubsection{Simulation with different models under multinomial
response} \label{sec:5.2.2}

In the second simulation the performance of the four models, used in
Section \ref{sec:5.1}, under a multinomial response with different
discontinuities is studied. In this simulation, time series with a
length of 11 time points are generated as follows. For each time point
$n_{t}$ independent trials are drawn from a multinomial distribution
with parameters $n_{t}$ and $\mathbf{p}_{t} =
(p_{t,1},p_{t,2},p_{t,3},p_{t,4})$, with $n_{t}$ the yearly sample size
and $\mathbf{p}_{t}$ the observed distribution over the four categories
of ``Contact frequency with neighbors'' observed with the PSLC in the
first 8 years and the SSPEC in the last 3 years. The distributions
observed with the SSPEC are corrected for the estimated discontinuities
obtained with model M2. Thus, $\mathbf{p}_{t} = \hat{\mathbf{y}}_{t}$ if
$t \le2004$ and $\mathbf{p}_{t} = \hat{\mathbf{y}}_{t} -
\hat{\bolds{\beta}}$ if $t > 2004$. According to this approach,
uninterrupted time series $\mathbf{p}_{t}^{r^{\ast}}$ are generated.

Subsequently, two different types of discontinuities are added to the
last three time points of the series, that is, $\mathbf{p}_{t}^{r} =
\mathbf{p}_{t}^{r^{\ast}} + \bolds{\Delta} _{t}$. The first set of
discontinuities are chosen constant over time by taking $\bolds{\Delta}
_{t} = (4.5, - 0.1, - 3.0, - 1.4)^{t}$ for $t = 2005, 2006$ and 2007.
These discontinuities are approximately equal to the estimated
discontinuities under model M2; see Table \ref{tab3}. The second set of
discontinuities is derived from the estimation results obtained with
model M3. Time varying discontinuities are obtained by taking
$\bolds{\Delta} _{t} = \hat{\mathbf{y}}_{t} - \tilde{\mathbf{y}}_{t}$
for $t = 2005, 2006$ and 2007. Here $\hat{\mathbf{y}}_{t}$ are the
originally observed series under the SSPEC and $\tilde{\mathbf{y}}_{t}$
the adjusted series obtained with the inverse of the logratio
transformation (\ref{eq14}). Although M3 assumes a time independent regression
coefficient for the intervention variable, the discontinuities become
time dependent since the adjusted series is mapped from the real space
back to the simplex with the inverse of the logratio transformation~(\ref{eq14}).

In each simulation 10,000 series are generated and analyzed with the
four models proposed in Section \ref{sec:5.1}. Let $\hat{\bolds{\Delta}}
{}^{r}_{t}$ denote the estimated discontinuities for time periods $t =
2005$, 2006 and 2007 for the $r$th replicate. For models M1 and M2 the
estimated discontinuities are equal to the estimated regression
coefficients of the intervention variable, that is,
$\hat{\bolds{\Delta}}{}^{r}_{t} = \hat{\bolds{\beta}}{}^{r}$, and thus
constant in time. For models~M3 and M4 the simulated series are
transformed using the logratio and the central logratio transformation
respectively. Time varying discontinuities for the $r$th replicate are
estimated as the difference between the original and adjusted series,
that is, $\hat{\bolds{\Delta}}{}^{r}_{t} = \mathbf{p}_{t}^{r} -
\tilde{\mathbf{p}}_{t}^{r}$, for $t = 2005, 2006$ and 2007. Here
$\tilde{\mathbf{p}}_{t}^{r}$ denotes the adjusted series for the $r$th
replicate obtained with the inverse of the logratio transformation (\ref{eq14})
or the inverse central logratio transformation (\ref{eq17}).

In Table \ref{tab5} the mean and standard errors of the estimated discontinuities
$\hat{\bolds{\Delta}}{}^{r}_{t}$ are summarized for the simulation with
constant discontinuities. Standard errors are obtained with the resample
standard deviation. In Table \ref{tab6} the same analysis results are specified
for the simulations with time dependent discontinuities. To compare the
simulation results of the models applied to the untransformed series
with the results obtained with the models applied to the transformed
series, the discontinuities estimated with models M3 and M4 are
transformed back to their original values on the simplex using the
approach described in the third paragraph of Section \ref{sec:5.2.2}.

For each model it follows that the difference between the real value and
the mean of the resample estimates of the discontinuities are small
compared to the standard errors, which implies that there are no
indications that one of the models results in biased parameter estimates
for the discontinuities. Nevertheless, it can be concluded that the
simulated means of the discontinuities of model M1 and M2 are closer to
the real values of the discontinuities than models M3 and M4. This is
the case for the simulation with constant discontinuities (Table \ref{tab5}) and
also for the time varying discontinuities (Table \ref{tab6}). Furthermore, the
simulated standard errors under models M1 and M2 are smaller than the
simulated standard errors obtained with models M3 and M4.

%
%t5 ###
\begin{table}
\caption{Real and simulated values time independent discontinuities}
\label{tab5}
\begin{tabular*}{\textwidth}{@{\extracolsep{\fill}}ld{1.12}ccc@{}}
\hline
% Row 1
& \multicolumn{4}{c@{}}{\textbf{Discontinuity}}\\[-6pt]
& \multicolumn{4}{c@{}}{\hrulefill}\\
% Row 2
& \multicolumn{1}{@{}c}{\hspace*{-6pt}\textbf{Cat. 1}} & \multicolumn{1}{c}{\textbf{Cat. 2}} & \multicolumn{1}{c}{\textbf{Cat. 3}} & \multicolumn{1}{c@{}}{\textbf{Cat. 4}} \\
% Row 3
\hline
Real value & 4.5 & $-$0.1\phantom{0000000.0} & $-$3.0\phantom{0000000.0} & $-$1.4\phantom{00000000} \\
% Row 4
M1 & 4.400\mbox{ }(1.232) & \phantom{0.}0.037\mbox{ }(0.631) & $-$2.672\mbox{ }(1.248) & $-$1.529\mbox{ }(0.489)\\
% Row 5
M2 & 4.266\mbox{ }(1.209) & $-$0.001\mbox{ }(0.650) & $-$2.694\mbox{ }(1.125) & $-$1.572\mbox{ }(0.497)\\
% Row 6
M3-2005 & 3.489\mbox{ }(1.430) & \phantom{0.}0.042\mbox{ }(0.759) & $-$1.818\mbox{ }(1.118) & $-$1.713\mbox{ }(0.578)\\
% Row 7
M3-2006 & 3.946\mbox{ }(1.682) & \phantom{0.}0.100\mbox{ }(0.685) & $-$2.274\mbox{ }(1.437) & $-$1.773\mbox{ }(0.696)\\
% Row 8
M3-2007 & 3.976\mbox{ }(1.677) & \phantom{0.}0.108\mbox{ }(0.745) & $-$2.038\mbox{ }(1.230) & $-$2.046\mbox{ }(0.850)\\
% Row 9
Mean value M3$^\ast$ & 3.804 & 0.083\phantom{00000} & $-$2.043\phantom{000000} & $-$1.844\phantom{000000} \\
% Row 10
M4-2005 & 3.353\mbox{ }(1.336) & \phantom{0.}0.191\mbox{ }(0.864) & $-$1.935\mbox{ }(1.443) & $-$1.609\mbox{ }(0.577)\\
% Row 11
M4-2006 & 3.852\mbox{ }(1.658) & \phantom{0.}0.230\mbox{ }(0.775) & $-$2.426\mbox{ }(1.853) & $-$1.657\mbox{ }(0.680)\\
% Row 12
M4-2007 & 3.847\mbox{ }(1.591) & \phantom{0.}0.256\mbox{ }(0.852) & $-$2.192\mbox{ }(1.707) & $-$1.911\mbox{ }(0.825)\\
% Row 13
Mean value M4$^\ast$ & 3.684  & 0.226\phantom{00000} & $-$2.184\phantom{000000} & $-$1.725\phantom{000000} \\
\hline
\end{tabular*}
\tabnotetext[]{}{$^\ast$: Mean over the three years. Standard errors between brackets.}
\end{table}

%
%t6 ###
\begin{table}
\caption{Real and simulated values time dependent discontinuities}
\label{tab6}
\begin{tabular*}{\textwidth}{@{\extracolsep{\fill}}ld{1.12}ccc@{}}
\hline
% Row 1
& \multicolumn{4}{c@{}}{\textbf{Discontinuity}}\\[-6pt]
& \multicolumn{4}{c@{}}{\hrulefill}\\
% Row 2
& \multicolumn{1}{c}{\hspace*{-12pt}\textbf{Cat. 1}} & \multicolumn{1}{c}{\textbf{Cat. 2}} & \multicolumn{1}{c}{\textbf{Cat. 3}} & \multicolumn{1}{c@{}}{\textbf{Cat. 4}} \\
% Row 3
\hline
% Row 3
Real value 2005 & 4 & $-$0.21\phantom{000000.0} & $-$1.96\phantom{000000.0} & $-$1.83\phantom{0000000} \\
% Row 4
Real value 2006 & 4.45 & $-$0.11\phantom{000000.0} & $-$2.46\phantom{000000.0} & $-$1.88\phantom{0000000}  \\
% Row 5
Real value 2007 & 4.47 & $-$0.12\phantom{000000.0} & $-$2.20\phantom{000000.0} & $-$2.15\phantom{0000000} \\
% Row 6
M1 & 3.788\mbox{ }(1.207) & $-$0.035\mbox{ }(0.614) & $-$1.562\mbox{ }(1.134) & $-$1.975\mbox{ }(0.446)\\
% Row 7
M2 & 3.665\mbox{ }(1.153) & $-$0.072\mbox{ }(0.629) & $-$1.582\mbox{ }(1.052) & $-$2.011\mbox{ }(0.459)\\
% Row 8
M3-2005 & 2.997\mbox{ }(1.245) & $-$0.041\mbox{ }(0.710) & $-$0.845\mbox{ }(0.932) &$-$2.111\mbox{ }(0.538)\\
% Row 9
M3-2006 & 3.207\mbox{ }(1.422) & $-$0.010\mbox{ }(0.645) & $-$0.993\mbox{ }(1.123) &$-$2.204\mbox{ }(0.681)\\
% Row 10
M3-2007 & 3.331\mbox{ }(1.461) & \phantom{0.}0.001\mbox{ }(0.703) & $-$0.896\mbox{ }(1.041) & $-$2.437\mbox{ }(0.830)\\
% Row 11
M4-2005 & 2.910\mbox{ }(1.153) & \phantom{0.}0.064\mbox{ }(0.781) & $-$0.925\mbox{ }(1.184) & $-$2.048\mbox{ }(0.548)\\
% Row 12
M4-2006 & 3.146\mbox{ }(1.361) & \phantom{0.}0.083\mbox{ }(0.705) & $-$1.095\mbox{ }(1.445) & $-$2.134\mbox{ }(0.679)\\
% Row 13
M4-2007 & 3.246\mbox{ }(1.348) & \phantom{0.}0.107\mbox{ }(0.774) & $-$0.996\mbox{ }(1.348) & $-$2.357\mbox{ }(0.813)\\
\hline
\end{tabular*}
\tabnotetext[]{}{Standard errors between brackets.}
\end{table}

%s5.3 ###
\subsection{Implementation} \label{sec:5.3}

The simulations indicate that time series models applied to the
untransformed series result in more accurate estimates for the
discontinuities than the models applied to the logratio or central
logratio transformed series. The main advantage of the logratio and
central logratio transformation is that the adjusted values add up to
one and always take values within the admissible range of $[0,1]$ by
definition. The major drawback of both transformations is that the
interpretation of the results is complex. The estimated discontinuities
as well as the corrected series for a particular class are influenced by
the discontinuity of the reference class in the case of the logratio
transformation. In the case of the central logratio transformation, the
estimated discontinuities as well as the corrected series for each
particular class are influenced by the discontinuities of all other
classes, via the geometric mean over all classes in the denominator of
this transformation. An additional disadvantage of the logratio
transformation is that the results depend on the choice of the reference
category to be used in the denominator of the logratio transformation.

The advantage of the multivariate model applied to the untransformed
data is that the interpretation of the results is straightforward and
that the estimated discontinuities for the separated categories are only
affected by the other categories through the zero sum constraint. The
major drawback is that the corrected values might take values outside
the admissible range of $[0,1]$. This, however, did not occur in this
application.

Based on these considerations, the multivariate model M2 applied to the
untransformed data is finally used in this application to estimate
discontinuities and calculate corrected time series for all other
parameters about environmental consciousness and social participation.
The common picture of the effect of the redesign is an increase of the
proportion of respondents in the first categories compensated by a
decrease in the last categories after the changeover. A more detailed
discussion about the results can be found in the supplemental paper, van den Brakel and Roels (\citeyear{vandenBrakelRoels09}).

In this application, the series for the two domains of gender were also
analyzed and adjusted for the observed discontinuities. For a few
parameters, the Lagrange function, described in Section \ref{sec:4.4}, was applied
to restore the consistency with the series for the total population. In
this case the covariance matrix in (\ref{eq20}) was taken diagonal with the
variances of the smoothed Kalman-filter estimates for the regression
coefficients of the intervention variables as elements. This benchmark
resulted in small modifications of the adjusted series.

Consistent time series can be obtained by correcting the observed series
for the estimated discontinuity. Depending on the anticipated impact of
the redesign on the quality of the estimates, the series observed in the
past can be adjusted to make it comparable with the outcomes obtained
under the new design. It is also possible to adjust the outcomes
obtained under the new approach to make them comparable with the series
under the old survey design. In this application the data collection
mode changed from CAPI under the PSLC to CATI under the SSPEC.
Therefore, it is anticipated that the series observed in the past are
more accurate than the outcomes obtained under the SSPEC. Indeed, with
the CAPI mode the entire target population is reached while the CATI
mode only surveys the subpopulation with a listed telephone number.
Furthermore, less measurement errors and socially desirable answers are
expected under the CAPI mode due to the personal contact with an
interviewer and the lower interview speed; see, for example,
Holbrook, Green and Krosnick (\citeyear{Holbrook03}) and Roberts (\citeyear{Roberts07}). Based on these considerations, it was
decided that the outcomes obtained under the SSPEC are corrected to make
the series comparable with the outcomes of the PSLC. Under the
assumption that the development observed with the CATI data is
representative for the entire target population, consistent time series
are obtained.

%s6 ###
\section{Discussion} \label{sec:6}

The relevance of official statistics, produced by national statistical
institutes, strongly depends on the comparability of the outcomes over
time. A redesign of the survey process generally results in
discontinuities in time series obtained with repeatedly conducted sample
surveys. To avoid the confounding of real developments with the
systematic effect induced by the redesign, structural time series models
with an intervention variable are developed to estimate the size of the
discontinuities. This approach relies on the assumption that there is no
structural change in the evolution of the series of the population value
at the moment that the survey is redesigned. Additional auxiliary
information and subject matter expert knowledge can be used to asses
whether the assumption that there is no structural change in the real
evolution of the population variable is tenable. Auxiliary time series
can be incorporated in the model to improve the estimates for the
discontinuities. If this assumption is questionable, experiments where
both surveys are run in parallel for some period of time should be
considered as an alternative.

The transition of the PSLC to the SSPEC resulted in systematic
differences in the estimates for parameters about environmental
consciousness and social participation. In this application, Gaussian
state-space models are applied to compositional time series which are
derived from variables with a multinomial response at each time period.
In a simulation study the performance of multivariate models applied to
untransformed, logratio transformed and central logratio transformed
series are compared. In this application the most accurate estimates for
the discontinuities are obtained with a multivariate model applied to
the untransformed series that accounts for the unit sum constraint. This
is a remarkable result, since the logratio and central logratio
transformations were considered to account for the multinomial response.
It is worthwhile to investigate to what extent simulation methods for
the analysis of non-Gaussian models further improve the accuracy of the
estimated discontinuities.

Another point of concern is the limited length of the available series.
Simulations indicate that the dispersion of the resample distribution of
the maximum likelihood estimates for the hyperparameters narrows rapidly
if the length of the available series increases. The dispersion of the
resample distribution of the smoothed estimates of the discontinuities,
on the other hand, remains more stable if the length of the series in
the simulations increases. Therefore, it appears that although the
maximum likelihood estimates of the hyperparameters of the state-space
models can be far from the true values under the available series, the
models already produce useful estimates for the discontinuities. This is
a plausible result. Most information about the size of the discontinuity
comes from the observations close to the moment of the survey redesign.
This also depends on the flexibility of the other model components. The
discontinuities are increasingly based on local observations close to
the moment of the survey redesign, as the trend and other model
components are more flexible.

One aspect of the time series approach is that more observations under
the new approach become available when time proceeds. The advantage is
that the discontinuities can be quantified more accurately if this
additional information becomes available. A concomitant drawback is that
the estimated discontinuities three years after redesigning the survey
are still subject to revisions. A publication policy is required to deal
with these revisions in practice. For this application it was decided to
base the final estimates for the discontinuities on the information
available up until 2007.

\section*{Acknowledgments}
The authors would like to thank the Associate editor and the referees
for a careful reading and for giving constructive comments on an earlier
draft of this paper.

\begin{supplement}
\stitle{Supplement}
\slink[doi]{10.1214/09-AOAS305SUPP}
\slink[url]{http://lib.stat.cmu.edu/aoas/305/supplement.zip}
\sdatatype{.zip}
\sdescription{The supplementary article contains
additional information about discontinuities in the target variables
about social participation and environmental consciousness that occurred
due to the changeover from the PSLC to the SSPEC. It contains a
description of the target variables about social participation and
environmental consciousness as well as an overview of the observed
differences that occurred during the year of the changeover from the
PSLC in 2004 to the SSPEC in 2005. Finally, the analysis results using
the time series model selected in Section \protect\ref{sec:5.3} are presented for these
variables. As an example, the estimated series and the corrected series
for three variables are provided.\\
\hspace*{12pt}This supplement also contains the Ox-program, used to conduct the
intervention analysis with the state-space models developed in this
paper. Input files (time series of ``contact frequency with neighbors''
and ``separating chemical waste'' and a series with the sample sizes of
the surveys for the different time points) are also provided to
illustrate the use of the program.}
\end{supplement}

\printaddresses

\end{document}